\documentclass[fleqn,usenatbib]{mnras}

\usepackage{newtxtext,newtxmath}

\usepackage[T1]{fontenc}

\usepackage{tipa}

\DeclareRobustCommand{\VAN}[3]{#2}
\let\VANthebibliography\thebibliography
\def\thebibliography{\DeclareRobustCommand{\VAN}[3]{##3}\VANthebibliography}

\usepackage{graphicx}	
\usepackage{amsmath}	
\usepackage{academicons} 
\usepackage{xcolor}
\usepackage{widetext}
\usepackage{url}
\usepackage[utf8]{inputenc}
\usepackage{hyperref}
\usepackage{orcidlink}
\usepackage{multicol}
\usepackage{pdflscape}

\newcommand{\kms}{\mbox{km s$^{-1}~$}} 
\newcommand{\kmse}{\mbox{km s$^{-1}$}}

\newcommand{\msun}{M$_{\odot}~$} 
\newcommand{\msune}{M$_{\odot}$}

\newcommand{\dgr}{$^{\circ}~$}

\newcommand{\teff}{$\rm T_{eff}$~}
\newcommand{\teffe}{$\rm T_{eff}$}
\newcommand{\logg}{$\log{g}$~}
\newcommand{\logge}{$\log{g}$}

\newcommand{\metale}{[M/H]}

\newcommand{\masyre}{mas yr$^{-1}$}
\newcommand{\name}{Sutlej~}
\newcommand{\namee}{Sutlej}

\title[Split Stellar Stream]{Discovery of a Split Stellar Stream In the Periphery of the Small Magellanic Cloud}

\author[Nidever et al.]{David L. Nidever\orcidlink{0000-0002-1793-3689}$^{1}$\thanks{E-mail: dnidever@montana.edu} \\
$^{1}$Department of Physics, Montana State University, P.O. Box 173840, Bozeman, MT 59717-3840\\
}

\date{Accepted XXX. Received YYY; in original form ZZZ}

\pubyear{2023}

\begin{document}
\label{firstpage}
\pagerange{\pageref{firstpage}--\pageref{lastpage}}
\maketitle

\begin{abstract}
I report the discovery of a stellar stream (Sutlej) using {\it Gaia} DR3 proper motions and XP metallicities located $\sim$15\dgr north of the Small Magellanic Cloud (SMC). The stream is composed of two parallel linear components (``branches'') approximately $\sim$8\dgr $\times$ 0.6\dgr in size and separated by 2.5\degr.
The stars have a mean proper motion of ($\mu_{\rm RA}$,$\mu_{\rm DEC}$)=($+$0.08 \masyre,$-$1.41 \masyre) which is quite similar to the proper motion of stars on the western side of the SMC.
The color magnitude diagram of the stream stars has a clear red giant branch, horizontal branch, and main sequence turnoff that is well-matched by a PARSEC isochrone of 10 Gyr, [Fe/H]=$-1.8$ at 32 kpc and a total stellar mass of $\sim$33,000 \msune.  The stream is spread out over an area of 9.6 square degrees and has a surface brightness of 32.5 mag arcsec$^{-2}$.
The metallicity of the stream stars from {\it Gaia} XP spectra extend over $-2.5$ $\leq$ [M/H] $\leq$ $-1.0$ with a median of [M/H]$=-1.8$.
The tangential velocity of the stream stars is 214 \kms compared to the values of 448 \kms for the Large Magellanic Cloud and 428 \kms for the SMC.  While the radial velocity of the stream is not yet known, a comparison of the space velocities using a range of assumed radial velocities, shows that the stream is unlikely to be associated with the Magellanic Clouds.
The tangential velocity vector is misaligned with the stream by $\sim$25\dgr which might indicate an important gravitational influence from the nearby Magellanic Clouds.
\end{abstract}


\section{Introduction}
\label{sec:intro}

According to the currently-favored hierarchical galaxy formation paradigm \citep[e.g.,][]{Peebles1965,Press1974}, galaxies started small and grew through merger events and accretion of smaller systems most of which were tidally stripped apart.  Starting over three decades ago, mounting evidence of stellar streams in the Milky Way's (MW) halo were discovered, with the Sagittarius stream \citep[e.g.,][]{Ibata2001,Newberg2002} being the most prominent with its two tidal tails wrapping around the MW.  With the advent of deep, wide-field, multi-band photometric surveys, the number of discovered stellar streams rose quickly with the Sloan Digital Sky Survey \citep[SDSS; ][]{York2000} leading the way with the ``Field of Streams'' that included the Orphan stream, Anticenter stream, and others \citep{Belokurov2006}. Some of the most impressive streams are those produced by disrupted globular clusters that are extremely thin but can stretch over many tens of degrees \citep[e.g.,][]{Grillmair2006a,Grillmair2006b}.  
See \citet{Newberg2016} for a more detailed review of stellar streams.

Not only are observed stellar streams a striking confirmation of the violent origin of galaxies through mergers and accretion events, but stellar streams can be used as tracers to probe the Galaxy's mass and constrain the \mbox{3-D} structure of the gravitational potential \citep[e.g.,][]{Johnston2005,Koposov2010}.  One of the most effective search algorithms is the ``matched-filter'' method that selects all stars lying close to an old isochrone in the color-magnitude diagram (CMD) at a certain distance.  A range of distances are search and the resulting on-sky stellar density plots inspected for linear features.  Often the filters are heavily weighted towards the blue, main-sequence turnoff portion of the isochrone which has a large number of stars compared to the MW foreground.


Until recently, most stellar stream work was confined to the northern hemisphere due to the predominance of large surveys like SDSS, PS1 \citep[Pan-STARRS 1,][]{Chambers2016} and ATLAS \citep{Tonry2018} that covered that region of the sky.  However, with the advent of the Dark Energy Camera \citep[DECam;][]{Flaugher2015}, the situation changed dramatically.
Using the deep, multi-band DES photometric catalog \citep{DES}, \citet{shipp2018} discovered 11 new stellar streams in a southern ``tour-de-force'' much like the northern SDSS ``Field of Streams''.
The DECam Local Volume Exploration Survey \citep[DELVE;][]{DrlicaWagner2021} is systematically covering the entire southern sky with DECam to search for dwarf galaxies and stellar streams and was recently used to discover the Jet stream \citep{Ferguson2022}.

Even though deep, multi-band photometry has been the mainstay of stellar stream searches for decades, other techniques can also be extremely effective.  The second data release of 
{\it Gaia} \citep[DR2;][]{Brown2018} produced precise proper motions for over a billion stars.  This allowed for a kinematic selection of stellar streams.  \citet{Ibata2019} used a new systematic search method \citep[\texttt{STREAMFINDER};][]{Malhan2018} that takes advantage of the kinematics to discover eight new stellar streams throughout the MW, including in the MW mid-plane, which has historically been avoided by stream searches due to the high number of MW disk stars that can confuse search algorithms and generate many false positives.

\begin{figure}
    \centering
    \includegraphics[width=0.48\textwidth]{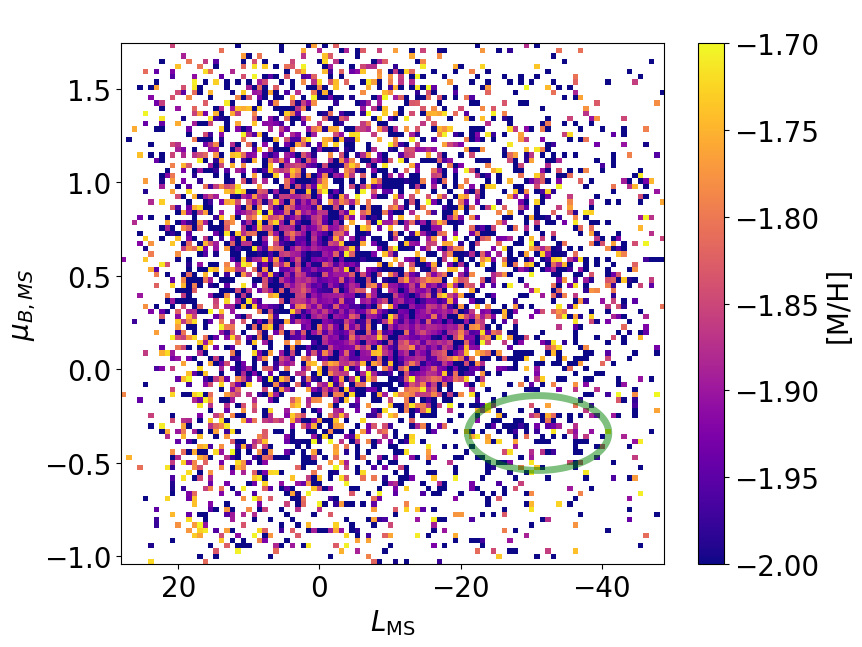}
    \caption{Density of Magellanic {\it Gaia} DR3 XP stars with [M/H]$\leq$$-1.7$ in the $\mu_{\rm B,MS}$ versus $L_{\rm MS}$.  The overdensity of stars with $-35$\dgr $\leq L_{\rm MS} \leq -20$\dgr and $\mu_{\rm B,MS}$ $\approx$ $-0.3$ mas yr$^{-1}$ is highlighted with the green ellipse.}
    \label{fig:mlonpmmb}
\end{figure}

In the third data release of {\it Gaia} \citep[DR3;][]{Vallenari2022}, the low-resolution {$BP$}/{$RP$} (XP) spectra of 220 million stars were released.  While the released stellar parameters \citep{Andrae2022} were not as reliable as originally anticipated, \citet{Andrae2023} determined precise metallicities as well as effective temperatures and surface gravities for 175 million stars using XGBoost trained on APOGEE \citep{Majewski2017} spectra and AllWise photometry \citep{Cutri2014}.

In this paper, I report on the discovery of a new stellar stream near the Small Magellanic Cloud using the {\it Gaia} DR3 proper motions and XP metallicities.

This paper is structured as follows. Section \ref{sec:data} discusses the data and catalogs while Section \ref{sec:methods} outlines the discovery and characterizes the main stream properties.
The main results are presented in Section \ref{sec:results} and the implications of the results are discussed in Section \ref{sec:discussion}.  Finally, the main conclusions are summarized in Section \ref{sec:summary}.

\section{Data}
\label{sec:data}

For this project, I used solely the {\it Gaia} DR3 dataset.  DR3 contains astrometric information, including proper motions, for 1.46 billion sources and three-band photometry ($G$, ${BP}$, and ${RP}$) for 1.54 billion sources.  In addition, it contains object classification from the ${BP}/{RP}$ (XP) spectra for 470 million sources, although the stellar parameters (\teffe, \logge, and \metale) in the official data release have some systematics that make the values unsuitable for most scientific analyses.  Instead, I use the stellar parameters from the \citet{Andrae2023} catalog that used a machine-learning model \citep[XGBoost;][]{Chen2016} trained on APOGEE \citep{Majewski2017,DR17} and other data to derive primarily \metale, but also \teff and \logg along the way, for 175 million stars from the average XP spectra that were released in {\it Gaia} DR3.
Note that \citet{Zhang2023} released a similar catalog but used a generative model to derive \teffe, \logge, \metale, and extinction for 220 million stars from the XP spectra.



\begin{figure}
    \centering
    \includegraphics[width=0.5\textwidth]{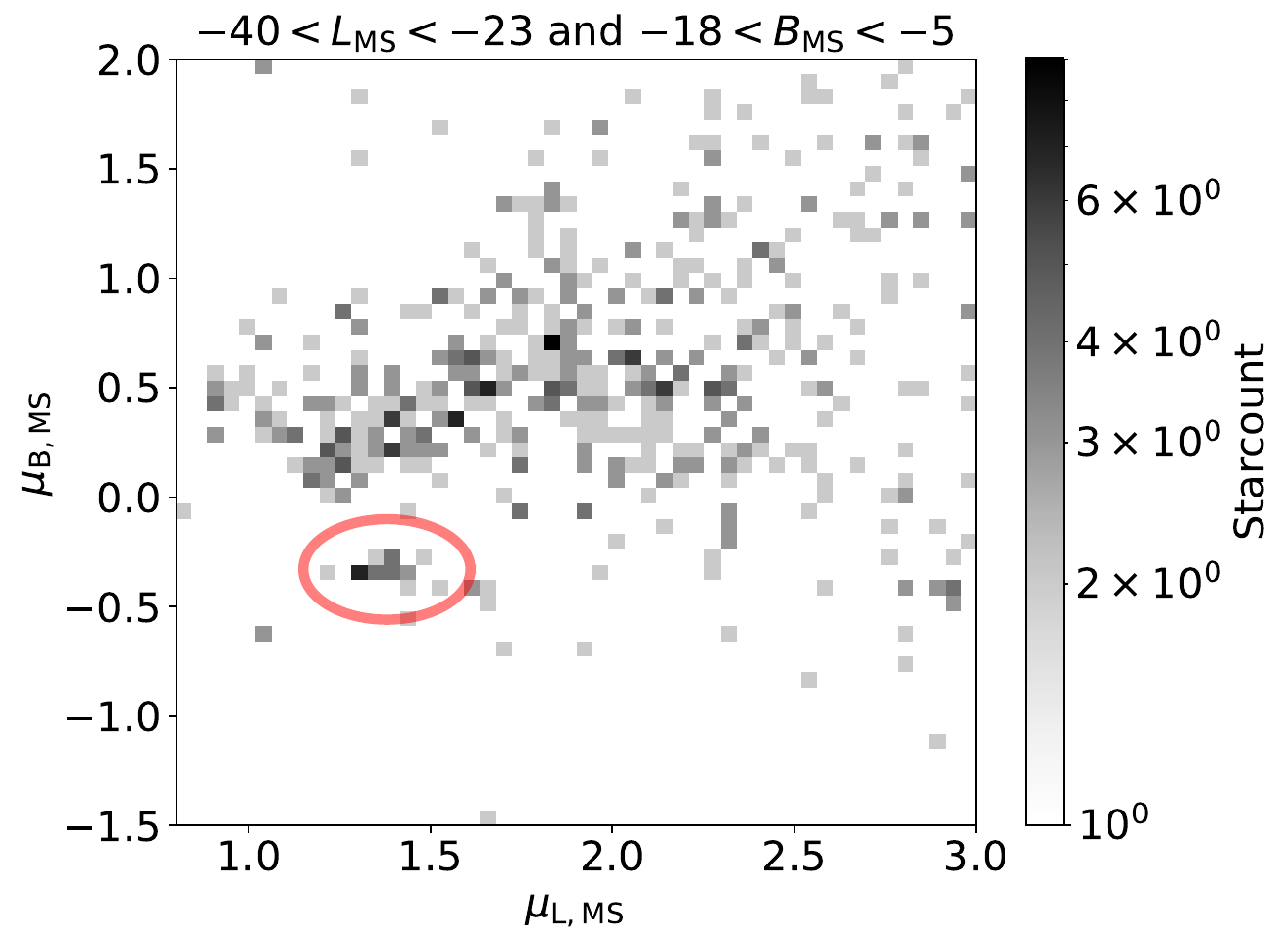}
    \caption{Density of Magellanic giant stars in $\mu_{\rm L,MS}$ versus $\mu_{\rm B,MS}$.  The stream stars are indicated by the red ellipse.}
    \label{fig:pmselection}
\end{figure}

\begin{figure*}
    \centering
    \includegraphics[width=0.9\textwidth]{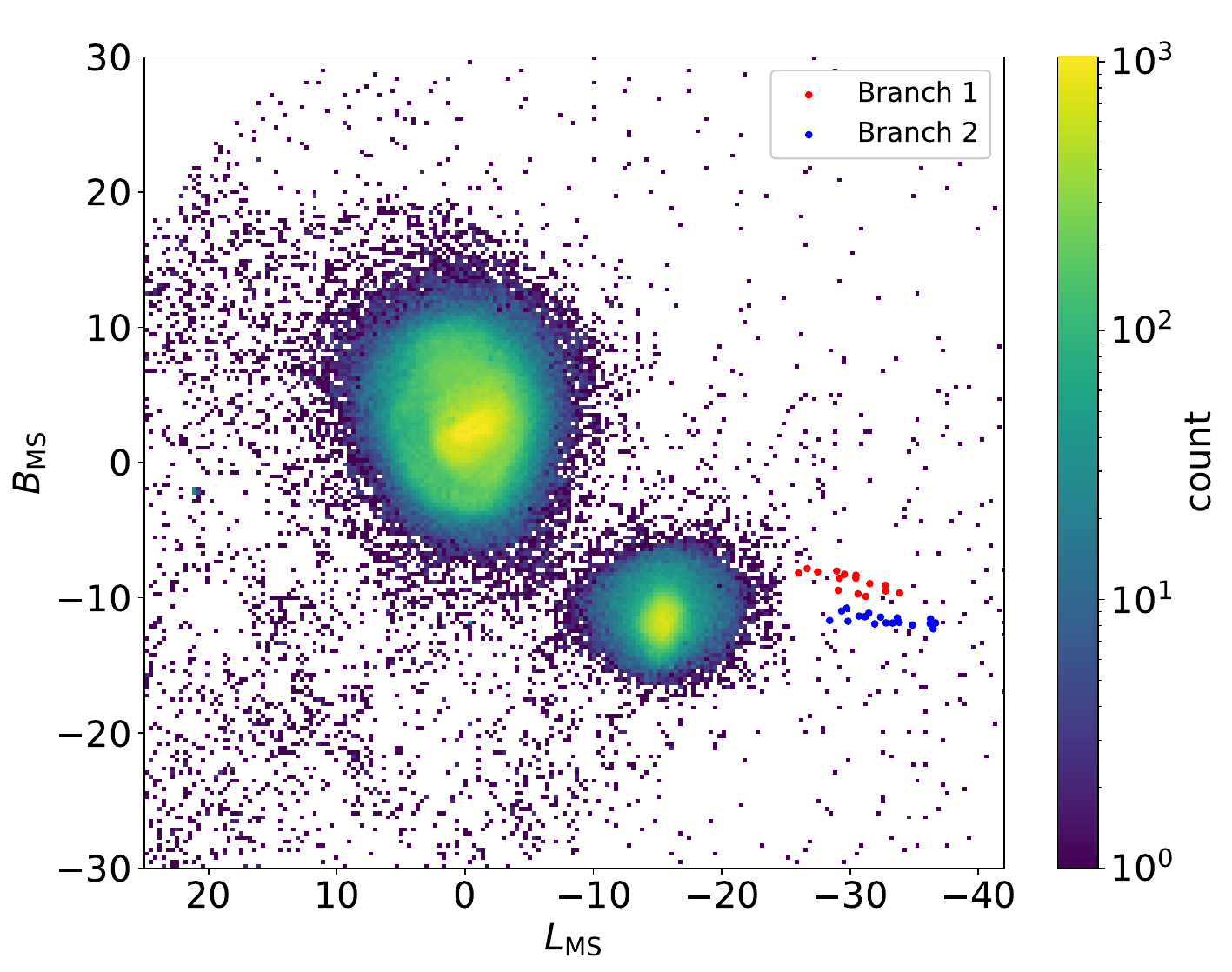}
    \caption{Density of Magellanic giant stars as selected from {\it Gaia} DR3 XP.  The position of the two stellar stream branches are indicated by the red (Branch 1) and blue (Branch 2) filled circles.}
    \label{fig:map}
\end{figure*}

\section{Discovery and Characterization}
\label{sec:methods}

While investigating the {\it Gaia} DR3 data for substructure in the periphery of the Magellanic Clouds (MCs) and looking through a variety of longitude versus proper motion figures in narrow ranges of metallicity, I serendipitously discovered an overdensity of stars.
Figure \ref{fig:mlonpmmb} shows an example of the type of figure that I was inspecting.  It shows the distribution of stars in the Andrae et al.\ catalog with \metale$<-$1.7 in the MC region in the $\mu_{\rm B,MS}$ vs.\ $L_{\rm MS}$\footnote{Where $L_{\rm MS}$/$B_{\rm MS}$ refer to the Magellanic Stream coordinate system defined in \citet{Nidever2008}, and $\mu_{\rm L,MS}$/$\mu_{\rm B,MS}$ are the corresponding proper motions.} color-coded by the metallicity. While the figure is quite ``messy'', a linear feature is visible in the bottom left-hand corner highlighted by the green ellipse ($-40$\dgr $\lesssim$ $L_{\rm MS}$ $\lesssim$ $-20$\degr, $\mu_{\rm B,MS}$ $\approx$ $-0.3$ \masyre). Further inspection showed that this feature is not spurious, but rather a real structure elongated spatially not far from the Small Magellanic Cloud (SMC).

Figure \ref{fig:pmselection} shows the proper motion distribution of the stars in that spatial region ($-40$\dgr $<$ $L_{\rm MS}$ $<$ $-23$\degr, $-18$\dgr $<$ $B_{\rm MS}$ $<$ $-5$\degr) exhibiting an even stronger overdensity at ($\mu_{\rm L,MS}$, $\mu_{\rm B,MS}$) = ($+$1.4 \masyre,$-1.3$ \masyre).  The stars selected by the red ellipse are plotted on the sky in Figure \ref{fig:map} as red and blue filled circles.  The background image shows the density of MC stars selected via proper motion, XP \teffe/\logge, and a red giant branch (RGB) box in the color magnitude diagram (CMD).  The new stellar structure is composed of two nearly-parallel stellar streams (``branches'') elongated on the sky by roughly $\sim$8 $\times$ 0.5--1\dgr and separated from each other by $\sim$2.5\degr.  At the closest point, the stream is only $\sim$1.5\dgr from the SMC Northern Overdensity \citep[SMCNOD;][]{Pieres2017} and $\sim$11\dgr from the center of the SMC.  This obviously begs the question of whether the new stream is associated with the SMC, which I explore in depth in Section \ref{sec:discussion}.

\begin{figure}
    \centering
    \includegraphics[width=0.47\textwidth]{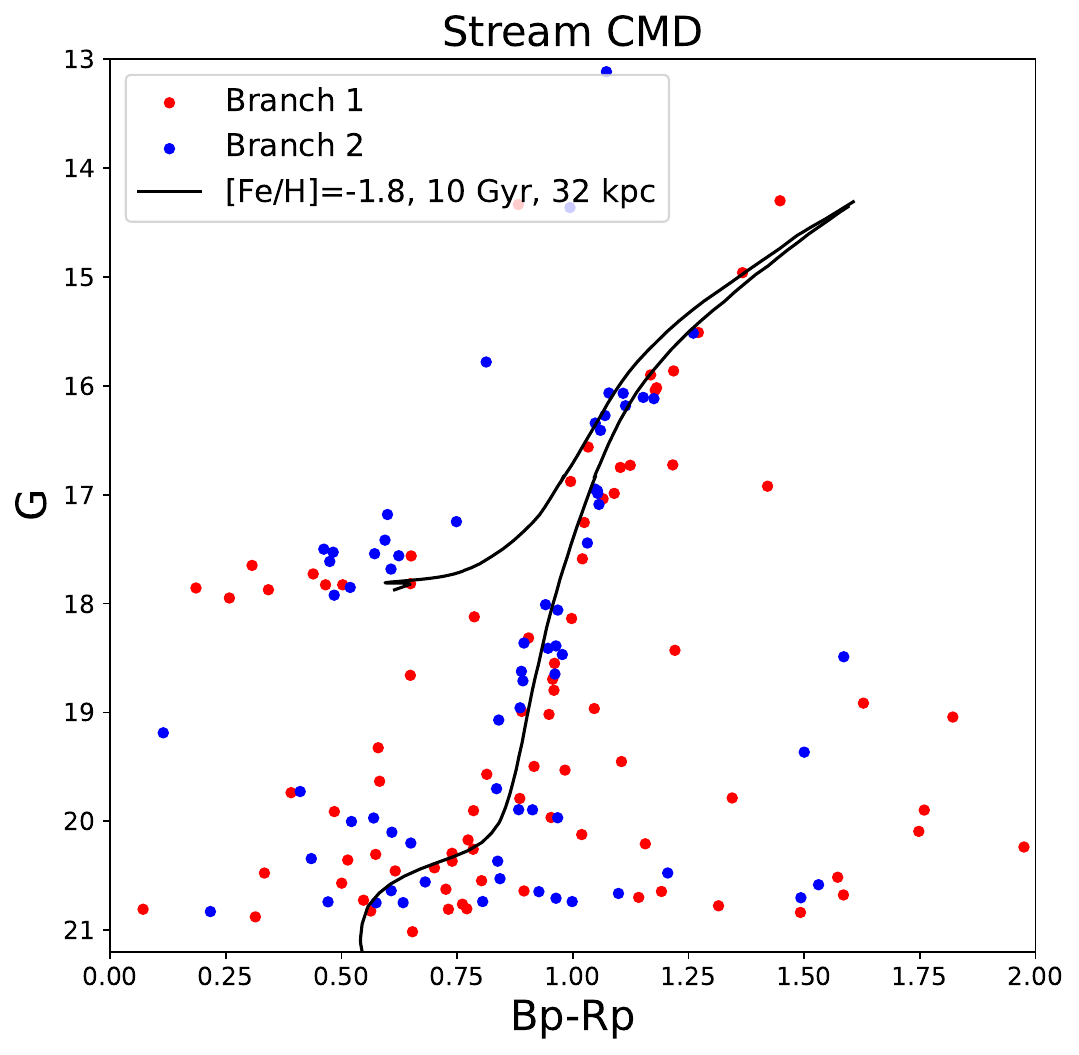}
    \caption{The color magnitude diagram of the region around the new stellar streams.  A clear red giant branch, horizontal branch, and main sequence turnoff are visible.  A PARSEC isochrone with [Fe/H]=$-1.8$, 10 Gyr and a distance of 32 kpc is shown in black.}
    \label{fig:gaiacmd}
\end{figure}

\begin{figure}
    \centering
    \includegraphics[width=0.47\textwidth]{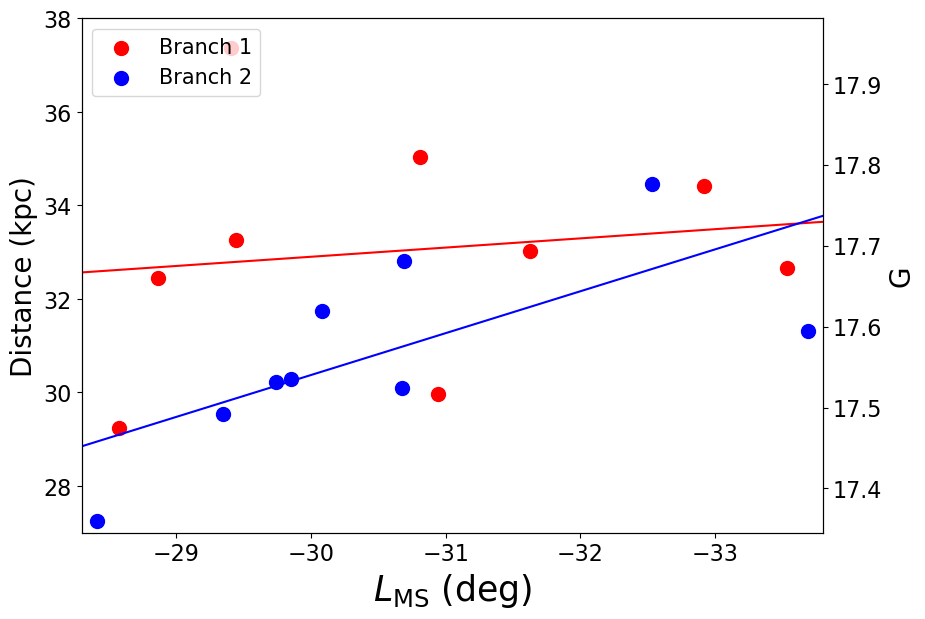}
    \caption{The distance (kpc) of the BHB stars versus $L_{\rm MS}$ for the two stream branches.  The BHB {\it Gaia} $G$ magnitude is shown on the right hand side.  The average BHB distance of Branch 1 is 33.0 kpc and for Branch 2 is 31.8 kpc. Branch 1 is $\sim$1.0 kpc farther away than Branch 2.  While Branch 1 does not show much of a distance gradient, the distance of the Branch 2 BHB stars grow larger with increasing angular distance from the SMC at a rate of $\sim$0.9 kpc per degree.}
    \label{fig:bhbdist}
\end{figure}

\begin{figure}
    \centering
    \includegraphics[width=0.47\textwidth]{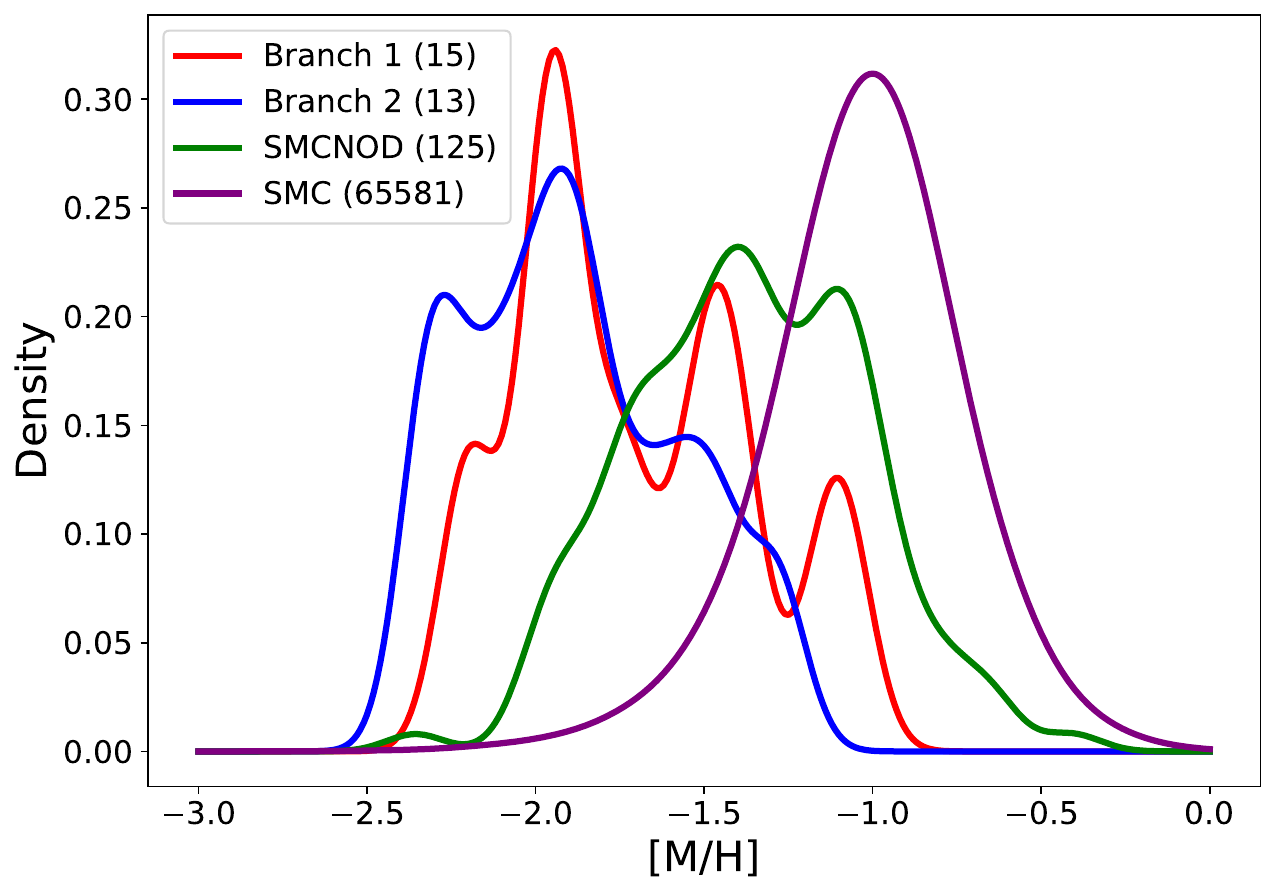}
    \caption{The metallicity distribution function (MDF) of the SMC (purple), the SMCNOD (green) and the stream stars (red and blue) using \citet{Andrae2023} {\it Gaia} DR3 XP metallicties.  The majority of the stream stars are more metal-poor than both the SMC (green) and SMCNOD (purple) stars with an average of [M/H]=$-1.65$ and peak around [M/H]=$-1.9$.}
    \label{fig:mdf}
\end{figure}

I will follow the naming convention of \citet{shipp2018} who named their stellar streams in this region of the sky after rivers in Pakistan and India, in particular after the Indus river and its tributaries Jhelum, Cheneb, and Ravi (in geographical order from northwest to southeast).  I shall name the new stream ``Sutlej'' (/\textipa{"s2tl@dZ}/) after the remaining main tributary of the Indus river which continues the geographical progression of the tributaries in northern India to the southeast and is nicely mirrored on the sky as the new stream is to the east of Ravi and Cheneb (see Fig.\ \ref{fig:streams}).



\begin{table*}
     \centering
     \caption{Table of coordinates of the stellar streams}
     \label{tab:summary}
     \begin{tabular}{ccccc}
     \hline
     Name & Equatorial Coordinates & Magellanic Stream Coordinates &
     Great Circle Pole & Width \\
    \hline
Branch 1 & [(3.2\degr,$-$55.2)\degr,(11.0\degr,$-$61.6\degr)] & [($-$33.9\degr,$-$9.7\degr),($-$26.4\degr,$-$8.0\degr)] & (68.7\degr, 16.3\degr) & 0.24\dgr \\
Branch 2 & [(357.9\degr,$-$53.2\degr),(2.9\degr,$-$60.3\degr)] & [($-$36.7\degr,$-$12.0\degr),($-$29.0\degr,$-$11.2\degr)] & (73.5\degr, 10.7\degr) & 0.29\dgr \\
    \hline
   \end{tabular}
\end{table*}

\section{Results}
\label{sec:results}

\begin{figure*}
    \centering
    \includegraphics[width=0.9\textwidth]{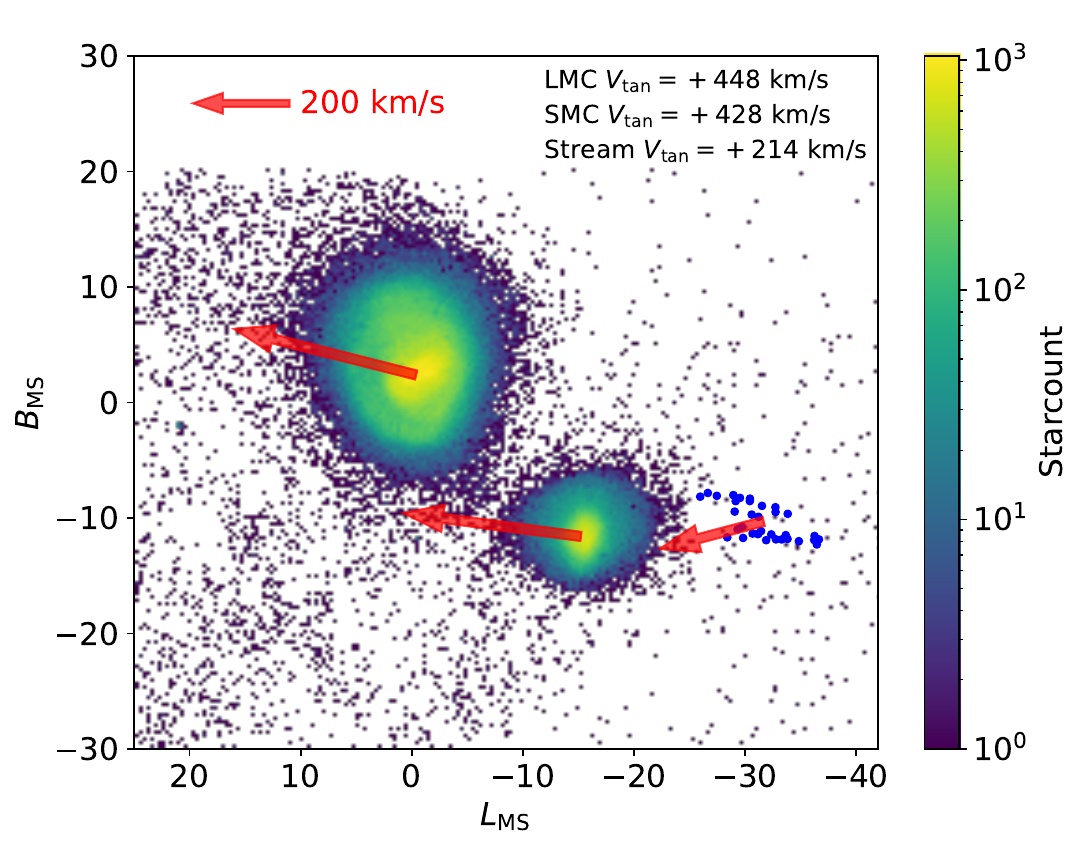}
    \caption{The tangential motion of the LMC, SMC and the stream stars.  The length of the red arrows indicates the magnitude of the velocity.  While \namee's tangential velocity vector points roughly in the direction of the MCs, they are different by $\sim$25\degr.  In addition, \namee's tangential velocity vector is misaligned with its stream track by $\sim$25\dgr indicating that it might have been gravitationally perturbed by the MCs.}
    \label{fig:maparrows}
\end{figure*}

The mean proper motion of the stream stars is $(\mu_{\rm L,MS},\mu_{\rm L,MS})$ = ($+1.4$ \masyre, $-0.322$ \masyre).
As can be seen in Figure \ref{fig:maparrows}, the tangential velocity vector is {\em not} aligned with the stream. In fact, it is $\sim$25\dgr misaligned.  This might be due to the influence of the nearly LMC and SMC, which has been shown to produce such a misalignment in the Orphan stream \citep{Erkal2019}.

Figure \ref{fig:gaiacmd} shows the color-magnitude diagram (CMD) of all {\it Gaia} DR3 stars in the vicinity of the two stream branches.  A well-defined red giant branch, blue horizontal branch (BHB), and main-sequence turnoff are apparent.  Both stellar branches are well-represented in these features with no large difference visible between them (except for the BHB; see below).  The black line shows a [Fe/H]=$-1.8$, 10 Gyr at 32 kpc PARSEC isochrone fit by-eye to the data.

While the CMDs of the two branches look nearly identical, Branch 1 (red) has a horizontal branch that extends 0.25 mag bluer than Branch 2 (blue) does.  The distances of the BHB stars can be directly estimated and contrasted with the isochrone distance and the distances of the two branches can be compared as well.  I converted the {\it Gaia} photometry to the SDSS system using the photometric transformation equations in the {\it Gaia} release documentation.\footnote{\url{https://gea.esac.esa.int/archive/documentation/GDR2/Data_processing/chap_cu5pho/sec_cu5pho_calibr/ssec_cu5pho_PhotTransf.html}}  The \citet{Barbosa2022} relation was used to compute the BHB absolute magnitude $M_g$ as a function of $g-r$ color and the distance calculated by comparing to the apparent magnitudes.  Figure \ref{fig:bhbdist} shows the BHB distances of the stream stars with a mean distance of 32.4 kpc for all stars and 33.0/31.8 kpc for Branch 1/2.  This is in excellent agreement with the isochrone-fitting distance of 32 kpc.  While Branch 1 shows almost no variation in distance with $L_{\rm MS}$, the Branch 2 distances increase the farther away they are from the SMC at a rate of $\sim$0.9 kpc per degree.

We can also investigate the metallicities of the brighter stream stars ($G<17.65$) using the {\it Gaia} DR3 XP metallicity provided by \citet{Andrae2023}.  Figure \ref{fig:mdf} shows the metallicity distribution function (MDF) of RGB stars from the stream (blue and red) as well as the SMC (purple) and SMCNOD (green).  To provide a smoother representation of the data (similar to a kernel density estimate), each star is presented as a Gaussian with unity amplitude and a FWHM of 0.2 dex.  The curve from each group is then divided by the number of stars to produce a density curve which makes the comparison between the stellar populations easier.  While the MDFs of the two stream branches look similar to each other (given the small number statistics), they are significantly more metal-poor than the majority of the SMC and SMCNOD distributions.  This alone is a strong indication that the stream did not originate from the SMC (but see \S \ref{sec:discussion} for more on the origin).  In addition, the broad stream MDF with a FWHM width of $\sim$1 dex suggests that the progenitor was a dwarf galaxy rather than a globular cluster.

I estimated the Great Circle Pole for each branch by searching a grid of pole coordinate values 0\degr $\leq$ $\alpha$ $\leq$ 360\degr, and 0 $\leq$ $\delta$ $\leq$ $+90$\dgr in steps of 1\dgr and calculated the {\it rms} (root-mean-square) of the transformed latitude for each pole.  I found the points coordinates with the lowest {\it rms} values and refined the search for more precision.
The pole for Branch 1 is ($\alpha$,$\delta$)=(68.7\degr,16.3\degr)
and for Branch 2 is 
($\alpha$,$\delta$)=(73.5\degr,10.7\degr).  The FWHM widths in latitude are 0.56\degr/0.68\dgr for Branch 1/2 which at a distance of 32 kpc corresponds to 0.31 kpc/0.38 kpc.  A summary of the stream values are shown in Table \ref{tab:summary}.


The total number of member stars in {\it Gaia} is low in the \name stream.  There are only 34 stream stars in the {\it Gaia} XP sample, and 80 in the full {\it Gaia} DR3 sample down to $G$=20.0 (RGB and BHB stars).  It is, therefore, worth estimating the stellar mass in the stream.  
I estimated the total stellar population mass by creating synthetic photometry from a 10 Gyr, [Fe/H]=$-1.8$ PARSEC isochrone at 32 kpc with a total stellar mass of $10^6$ \msune.  All RGB and BHB synthetic stars down to a magnitude of $G$=20.0 were selected and the same operation performed on the data.  This resulted in 80 {\it Gaia} stars and 6056 synthetic stars.  Scaling the input isochrone mass of $10^6$ \msun by 80/6056 gives 13,210 \msune.  However, by inspecting the two luminosity functions it became clear that they do that match well at the faint end; the observed number of stars does not increase as quickly as expected from the isochrone likely due to incompleteness.  A more realistic estimate of 33,333 \msun was found by finding the best scaling of the luminosity functions. 

Finally, I calculate the surface brightness by summing up the flux from all synthetic stars and scaling to the total mass of 33,333 \msune.  The area of each stream branch is 0.6\dgr $\times$ 8\dgr or 9.6 square degrees combined.  The total flux is divided by the area in arcsec squared and converted back to magnitude to obtain 32.3 mag arcsec$^2$.
Another way to calculate the surface brightness is to sum up the flux of the observed stars and then correct for incompleteness.  Using the same procedure with the 80 RGB and BHB stars down to $G$=20.0 gives a surface brightness of 33.1 mag arcsec$^2$.  However, this is incomplete because of the stars that we are not seeing.  We can use the theoretical isochrone to calculate a good estimate for this completeness by calculating the cumulative fraction of total flux as a function of $G$.  At 19 $\leq$ $G$ $\leq$ 20 mag this is fairly constant with a value of $\sim$55\%.  Applying this correction to the total observed flux gives a surface brightness of 32.5 mag arcsec$^2$ which is quite close to the 32.3 mag arcsec$^2$ calculated with the ischrone method above.

\begin{figure*}
    \centering
    \includegraphics[width=1.0\textwidth]{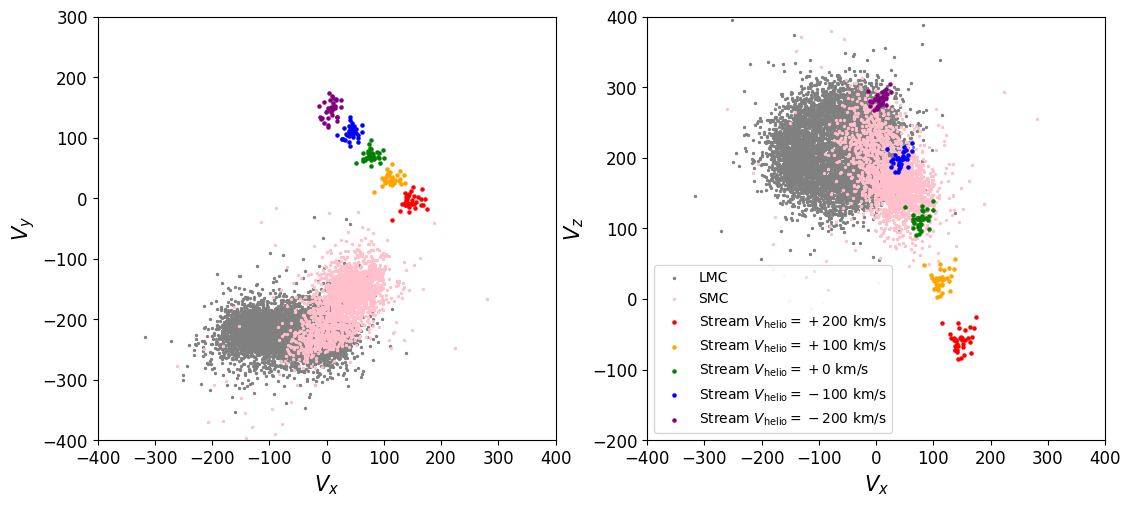}
    \caption{The galactocentric space velocities of the LMC (gray), SMC (pink) and the \name stream stars (colored points).  The colored points show the space velocities of the stream stars assuming a range of radial velocity: $-200$ \kms (purple), $-100$ \kms (red), 0 \kms (green), $+100$ \kms (orange), and $+200$ \kms (red).  For all radial velocities, the space velocities of the stream stars are significantly offset from the LMC and SMC making a Magellanic origin unlikely.}
    \label{fig:spacevelocity}
\end{figure*}


\section{Discussion}
\label{sec:discussion}

As previously mentioned, an obvious question, due to its proximity, is whether \name is related to the Magellanic Clouds.  The edge of the stream is only a couple of degrees away from the SMCNOD and the two stream branches are elongated almost parallel to the LMC's and SMC's tangential velocity vectors (see Figure \ref{fig:maparrows}).  In addition, the stream distance of 32 kpc is not too dissimilar to the Magellanic Clouds' distance of 50 kpc (LMC) and 60 (SMC).  This is, however, where the similarities end.  As the MDF shows (Fig. \ref{fig:mdf}), the stream stars are much more metal-poor than the MCs metallicity distributions (LMC is $\sim$0.4 dex more metal-rich than the SMC).  The tangential velocity vector is also misaligned with the LMC/SMC's by $\sim$25\dgr (Fig.\ \ref{fig:maparrows}) and its total tangential velocity of 214 \kms is almost a factor of two smaller than the MCs ($\sim$430 \kmse).  And although we do not know \namee's radial velocity (RV) yet, we can compare space velocities with the MCs.
Figure \ref{fig:spacevelocity} shows the 3-D space velocities of the stream stars compared to the APOGEE MC stars assuming a wide range of stream RVs.  No matter the stream RV, the space velocities are always offset from the MC space velocities by more than $\gtrsim$200 \kmse.  Therefore, a Magellanic origin seems unlikely.

\begin{figure}
    \centering
    \includegraphics[width=0.47\textwidth]{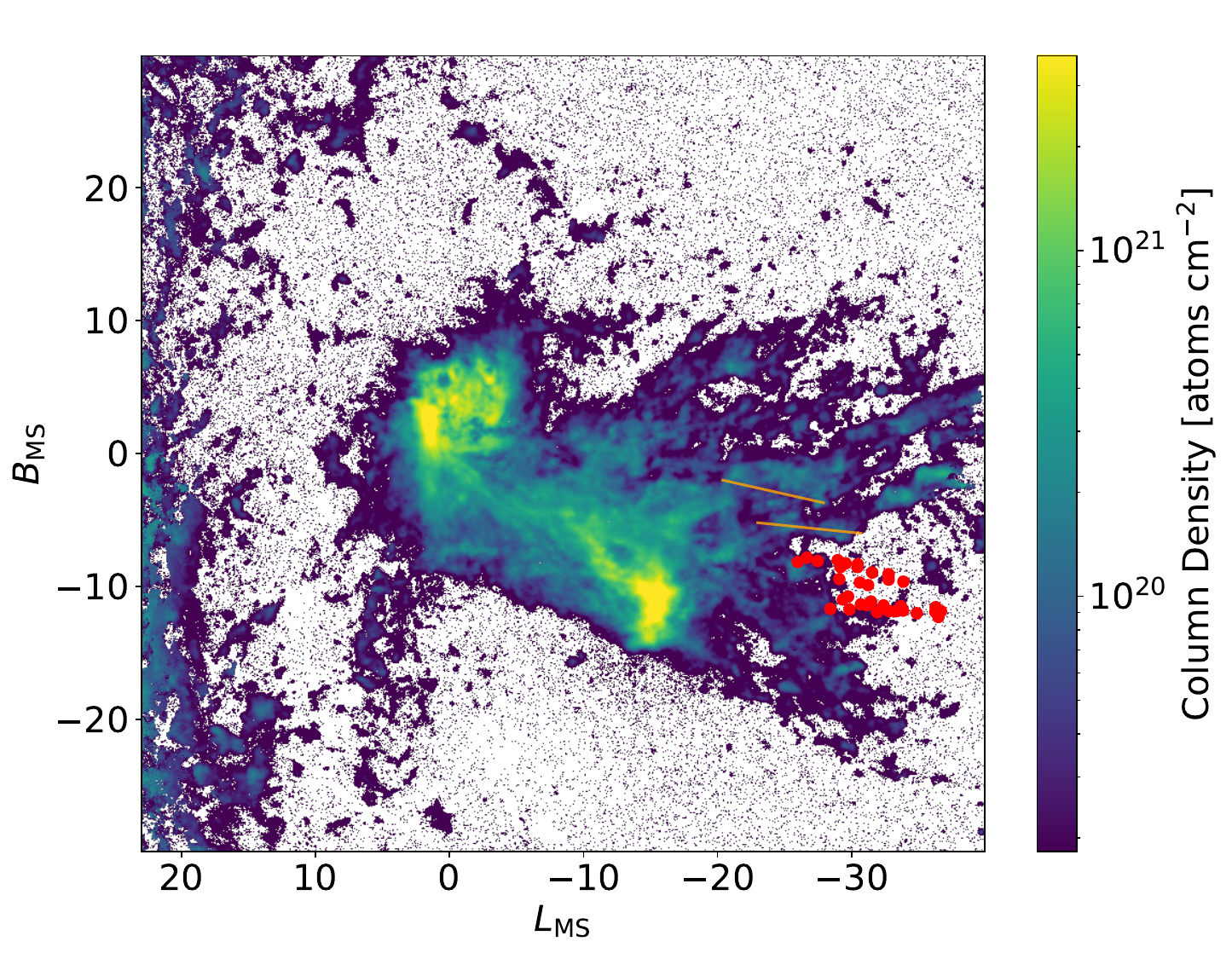}
    \caption{Column density map of HI from the GASS survey \citep{McClure-Griffiths2009}.  The new stellar stream is shown in filled red circles.  Orange lines indicate the new stellar stream offset by $+6$\dgr in $L_{\rm MS}$ and $+6$\dgr in $B_{\rm MS}$.}
    \label{fig:gass}
\end{figure}

However, I must point out one interesting spatial pattern in the two \name branches that reminds me of features in the two filaments of the Magellanic Stream (MS) that are quite nearby in the sky.  Figure \ref{fig:gass} shows the GASS \citep{McClure-Griffiths2009} HI column density map of the Magellanic System including the Stream and the \name stars (red filled circles).  The two MS filaments show structures that are roughly 2\dgr $\times$ 10\dgr in size and offset from each other by a few degrees.  This pattern looks very similar to the \name branches.  In fact, I have fit lines to the \name branches and offset them both by $+$6\dgr in $L_{\rm MS}$ and $+$6\dgr in $B_{\rm MS}$ (thin orange lines).  They match the length and the orientation of the MS features quite well, although one of them is offset by roughly a degree.  
Even if \name and these MS features were related, it does not seem realistic that the gas is {\it leading} the stars in their orbit.  The Price-Whelan 1 star cluster \citep{PriceWhelan2019,Nidever2019} was born in the Magellanic Stream Leading Arm 117 Myr ago and has since separated from and is leading the gas by $\sim$10\degr.  This is understandable and expected due to the ram pressure effects of the MW's hot halo gas on the Leading Arm gas.  In fact, this can be used to constrain the density of the hot halo (see \S4.5 of \citealt{Nidever2019}).  However, if the \name branches and the MS linear features are somehow causally related, then it is unlikely that the stars would be trailing the gas.  In addition, figuring in all of the discrepancies of an MC origin of \name mentioned above, this linear feature in \name and the MS is likely a curious coincidence.

\begin{figure}
    \centering
    \includegraphics[width=0.47\textwidth]{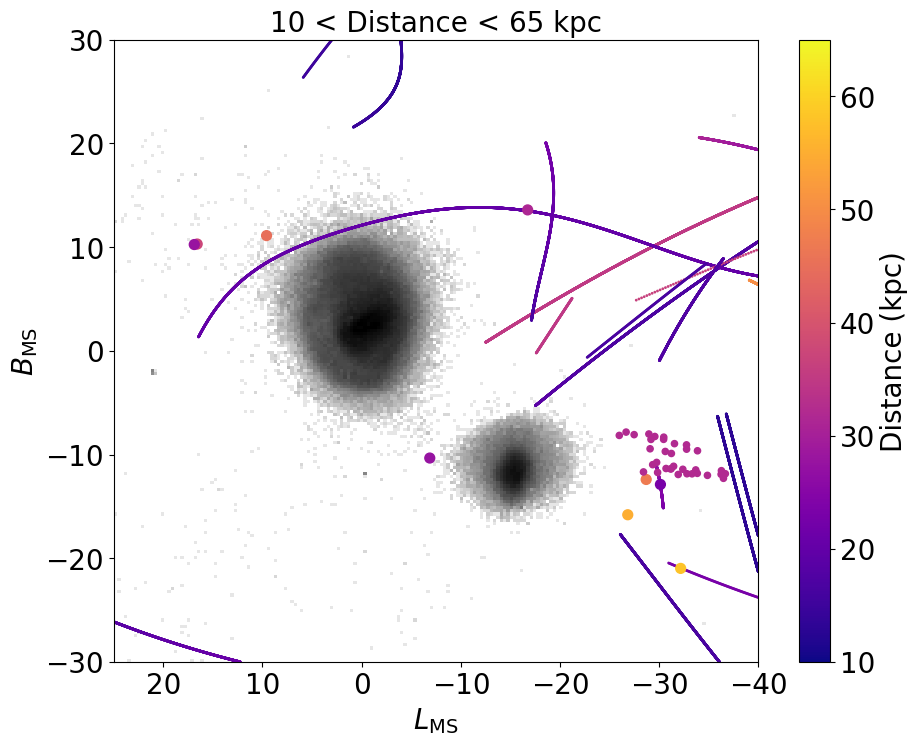}
    \caption{The density of Magellanic giant stars from {\it Gaia} DR3 with the known stellar streams \citep{Mateu2023} and 
      dwarf spheroidal galaxies between 10 to 65 kpc away color-coded by their distance.
     The new stellar stream is indicated by purpled filled circles}
    \label{fig:streams}
\end{figure}

Could the \name stream be related to any of the streams or dwarf galaxies discovered in the MC region by DES, DELVE \citep{DrlicaWagner2021} and others?  Figure \ref{fig:streams} shows the streams (as tabulated by \citealt{Mateu2023}) and dwarf galaxies near the MCs with distances of 10 to 65 kpc color-coded by distance.  None of the nearby stellar streams are aligned with \namee.  Moreover, although there are dwarf galaxies nearby they are either at larger distances (Phoenix II at 84 kpc, \citealt{Bechtol2015}; Tucana IV at 47 kpc and Tucana V at 55 kpc, \citealt{Drlica-Wagner2015}) or are moving in a different direction \citep[Tucana III;][]{Drlica-Wagner2015}.  The only exception is Hydrus 1 \citep{Koposov2018} which is at a distance of 28 kpc and on the opposite side of the SMC from \name but close to an extrapolation of the \name branch tracks.
Selecting Gaia DR3 stars near Hydrus 1, I was able to determine a significant overdensity of 17 stars in proper motion space corresponding to ($\mu_{\rm L,MS}$,$\mu_{\rm B,MS}$)=($+$3.787 \masyre, $+$1.68 \masyre).  The CMD of these stars indicate that these are the Hydrus 1 RGB and BHB stars.  Using the $V_{\rm helio}$ = $+$80.4 \kms and 28 kpc distance from \citet{Koposov2018}, we can calculate the galactocentric space velocity of Hydrus 1 to be ($V_x$,$V_y$,$V_z$) = ($-$435.0 \kmse, $-$82.8 \kmse, $-$11.1 \kmse).  Comparing these values to Figure \ref{fig:spacevelocity}, it is clear that Hydrus 1 has a significantly different space velocity from the LMC, SMC and \name by $\gtrsim$200 \kmse.
In addition, while they are both metal-poor, Hydrus 1 is more metal-poor ([Fe/H]$=-2.5$) than \name ([Fe/H]$=-1.9$) and comparing the \name MDF in Figure \ref{fig:mdf} with the Hydrus 1 MDF in Figure 19 from \citet{Koposov2018} indicates that the metallicity distributions are quite different.

Another curious feature of \name is the two parallel split branches which is quite uncommon in stellar streams.  Some other examples are the Sagittarius stream \citep[e.g.,][]{Majewski2003,Koposov2012} that has bifurcated branches both in the leading and trailing arms and the Anticenter/Monoceros stream \citep[e.g.,][]{Grillmair2006e}.  The Sagittarius stream wraps around the MW due to the multiple passages its host galaxy made, and it is quite likely that the multiple pericentric passages created the bifurcation of the two tidal arms.  After much debate over the origin of Monoceros \citep{Martin2006,Conn2008}, it is now thought that this broad feature was produced by a perturbation of the outer MW disk by a large satellite like Sagittarius dwarf spheroidal galaxy or the LMC \citep{Slater2014,Morganson2016,Hayes2018b}. 
Since the two \name branches have nearly identical metallicity, distance, age, and space velocity, it seems quite likely that they have the same progenitor galaxy.  How exactly a small dwarf spheroidal galaxy could produce two parallal stellar streams offset by $\sim$2.5\dgr or 1.4 kpc (at 32 kpc) perpendicular to their orbit remains unclear.  However, this feature of \name should put tight constraints on simulations trying to reproduce it.

\citet{Erkal2019} used the Orphan stream to constrain the LMC's total mass to 1.38 $\times$ 10$^{11}$ \msun by using the perturbations that the LMC made on the Orphan stream's proper motions which are misaligned with the stream's track.  This is the most accurate mass estimate of the LMC to date.  The SMC's mass, on the other hand, is not well constrained and it is not unusual to scale the LMC's total mass by the ratio of the SMC and LMC's stellar masses ($\sim$1/9.6) as done by \citet{Besla2012}.  Having a more accurate SMC total mass would help improve simulations of the Magellanic system.  The \name stream provides a tantalizing possibility of being able to constrain the SMC's total mass since the \name proper motion vector is misaligned with its stream track.  Orbit modeling of the Gaia kinematics and follow-up spectroscopic radial velocities should be able to determine if this feat is possible. 

\section{Summary}
\label{sec:summary}

I report the discovery of a split stellar stream, 
\namee, near the Small Magellanic Cloud using {\it Gaia} DR3 proper motions and metallicities from the low-resolution XP spectra.  The main conclusions are:

\begin{itemize}
\item \name has two nearly-parallel branches that are roughly $\sim$8\dgr $\times$ 0.6\dgr in shape and separated by $\sim$2.5\degr.  They are situated $\sim$15\dgr north of the SMC.
\item The {\it Gaia} CMD shows a clear signature of a simple stellar population (with RGB, BHB and main-sequence turnoff) that is well-fit with an isochrone of age 10 Gyr, [Fe/H]=$-1.8$ and distance of 32 kpc.  
\item \name has a prominent blue horizontal branch.  Measured distances of these standard candles give a mean distance of  32.4 kpc for all \name stars and 33.0/31.8 kpc for Branch 1/2.  While Branch 1 shows little distance variation, Branch 2 has a distance gradient of $\sim$0.9 kpc deg$^{-1}$ where the distance increases as the angular distance from the SMC grows larger.
\item The {\it Gaia} XP metallicites show that \name has a broad MDF stretching from [Fe/H]$=-$2.5 to $-$1.0 with a median of [Fe/H]$=-1.9$.  The broad MDF strongly suggests that the progenitor was a dwarf galaxy rather than a globular cluster.
\item The total stellar mass of \name is 33,333 \msun and its surface brightness is 32.5 mag arcsec$^{-2}$.
\item \namee's tangential velocity vector is misaligned with its stream track by $\sim$25\dgr providing evidence that it has likely been gravitational perturbed by the nearby MCs.
\item \name is likely not associated with the SMC because no matter what the radial velocity is, the 3-D space velocity of \name is significantly offset from the SMC by at least $\sim$200 \kmse.  In addition, \namee's MDF is much more metal-poor than the MCs' and the tangential velocity vectors are misaligned by $\sim$25\degr.
\end{itemize}

Follow-up spectroscopic observations will to measure the radial velocity and chemical abundances should help resolve the origin of \name.
Orbital modeling of \name has the potential of constraining the mass of the SMc.

\section*{Acknowledgements}

I want to thank Andrew Pace for sharing with me his table of Milky Way dwarf galaxy properties.

This work has made use of data from the European Space Agency (ESA) mission
{\it Gaia} (\url{https://www.cosmos.esa.int/gaia}), processed by the {\it Gaia}
Data Processing and Analysis Consortium (DPAC,
\url{https://www.cosmos.esa.int/web/gaia/dpac/consortium}). Funding for the DPAC
has been provided by national institutions, in particular the institutions
participating in the {\it Gaia} Multilateral Agreement.

\textit{Software:} Astropy \citep{pricewhelan2018astropy, robitaille2013astropy}, Matplotlib \citep{hunter2007matplotlib}, NumPy \citep{harris2020numpy}, SciPy \citep{virtanen2020scipy}

\section*{Data Availability}


All {\it Gaia} DR3 data are available from the {\it Gaia} Archive \url{https://gea.esac.esa.int/archive/}.  The \citet{Andrae2023} catalog of {\it Gaia} DR3 XP stellar parameters and metallicities is available from \url{https://zenodo.org/record/7945154}.


\bibliographystyle{aasjournals}
\bibliography{ref_og.bib}

\begin{thebibliography}{}
\expandafter\ifx\csname natexlab\endcsname\relax\def\natexlab#1{#1}\fi
\providecommand{\url}[1]{\href{#1}{#1}}

\bibitem[{{Abdurro'uf} {et~al.}(2022){Abdurro'uf}, {Accetta}, {Aerts}, {Silva
  Aguirre}, {Ahumada}, {Ajgaonkar}, {Filiz Ak}, {Alam}, {Allende Prieto},
  {Almeida}, {Anders}, {Anderson}, {Andrews}, {Anguiano}, {Aquino-Ort{\'\i}z},
  {Arag{\'o}n-Salamanca}, {Argudo-Fern{\'a}ndez}, {Ata}, {Aubert},
  {Avila-Reese}, {Badenes}, {Barb{\'a}}, {Barger}, {Barrera-Ballesteros},
  {Beaton}, {Beers}, {Belfiore}, {Bender}, {Bernardi}, {Bershady}, {Beutler},
  {Bidin}, {Bird}, {Bizyaev}, {Blanc}, {Blanton}, {Boardman}, {Bolton},
  {Boquien}, {Borissova}, {Bovy}, {Brandt}, {Brown}, {Brownstein}, {Brusa},
  {Buchner}, {Bundy}, {Burchett}, {Bureau}, {Burgasser}, {Cabang}, {Campbell},
  {Cappellari}, {Carlberg}, {Wanderley}, {Carrera}, {Cash}, {Chen}, {Chen},
  {Cherinka}, {Chiappini}, {Choi}, {Chojnowski}, {Chung}, {Clerc}, {Cohen},
  {Comerford}, {Comparat}, {da Costa}, {Covey}, {Crane}, {Cruz-Gonzalez},
  {Culhane}, {Cunha}, {Dai}, {Damke}, {Darling}, {Davidson}, {Davies},
  {Dawson}, {De Lee}, {Diamond-Stanic}, {Cano-D{\'\i}az}, {S{\'a}nchez},
  {Donor}, {Duckworth}, {Dwelly}, {Eisenstein}, {Elsworth}, {Emsellem},
  {Eracleous}, {Escoffier}, {Fan}, {Farr}, {Feng}, {Fern{\'a}ndez-Trincado},
  {Feuillet}, {Filipp}, {Fillingham}, {Frinchaboy}, {Fromenteau}, {Galbany},
  {Garc{\'\i}a}, {Garc{\'\i}a-Hern{\'a}ndez}, {Ge}, {Geisler}, {Gelfand},
  {G{\'e}ron}, {Gibson}, {Goddy}, {Godoy-Rivera}, {Grabowski}, {Green},
  {Greener}, {Grier}, {Griffith}, {Guo}, {Guy}, {Hadjara}, {Harding},
  {Hasselquist}, {Hayes}, {Hearty}, {Hern{\'a}ndez}, {Hill}, {Hogg},
  {Holtzman}, {Horta}, {Hsieh}, {Hsu}, {Hsu}, {Huber}, {Huertas-Company},
  {Hutchinson}, {Hwang}, {Ibarra-Medel}, {Chitham}, {Ilha}, {Imig}, {Jaekle},
  {Jayasinghe}, {Ji}, {Johnson}, {Jones}, {J{\"o}nsson}, {Katkov}, {Khalatyan},
  {Kinemuchi}, {Kisku}, {Knapen}, {Kneib}, {Kollmeier}, {Kong}, {Kounkel},
  {Kreckel}, {Krishnarao}, {Lacerna}, {Lane}, {Langgin}, {Lavender}, {Law},
  {Lazarz}, {Leung}, {Leung}, {Lewis}, {Li}, {Li}, {Lian}, {Liang}, {Lin},
  {Lin}, {Lin}, {Lintott}, {Long}, {Longa-Pe{\~n}a}, {L{\'o}pez-Cob{\'a}},
  {Lu}, {Lundgren}, {Luo}, {Mackereth}, {de la Macorra}, {Mahadevan},
  {Majewski}, {Manchado}, {Mandeville}, {Maraston}, {Margalef-Bentabol},
  {Masseron}, {Masters}, {Mathur}, {McDermid}, {Mckay}, {Merloni},
  {Merrifield}, {Meszaros}, {Miglio}, {Di Mille}, {Minniti}, {Minsley},
  {Monachesi}, {Moon}, {Mosser}, {Mulchaey}, {Muna}, {Mu{\~n}oz}, {Myers},
  {Myers}, {Nadathur}, {Nair}, {Nandra}, {Neumann}, {Newman}, {Nidever},
  {Nikakhtar}, {Nitschelm}, {O'Connell}, {Garma-Oehmichen}, {Luan Souza de
  Oliveira}, {Olney}, {Oravetz}, {Ortigoza-Urdaneta}, {Osorio}, {Otter},
  {Pace}, {Padilla}, {Pan}, {Pan}, {Parikh}, {Parker}, {Peirani}, {Pe{\~n}a
  Ram{\'\i}rez}, {Penny}, {Percival}, {Perez-Fournon}, {Pinsonneault},
  {Poidevin}, {Poovelil}, {Price-Whelan}, {B{\'a}rbara de Andrade Queiroz},
  {Raddick}, {Ray}, {Rembold}, {Riddle}, {Riffel}, {Riffel}, {Rix}, {Robin},
  {Rodr{\'\i}guez-Puebla}, {Roman-Lopes}, {Rom{\'a}n-Z{\'u}{\~n}iga}, {Rose},
  {Ross}, {Rossi}, {Rubin}, {Salvato}, {S{\'a}nchez}, {S{\'a}nchez-Gallego},
  {Sanderson}, {Santana Rojas}, {Sarceno}, {Sarmiento}, {Sayres}, {Sazonova},
  {Schaefer}, {Schiavon}, {Schlegel}, {Schneider}, {Schultheis}, {Schwope},
  {Serenelli}, {Serna}, {Shao}, {Shapiro}, {Sharma}, {Shen}, {Shetrone}, {Shu},
  {Simon}, {Skrutskie}, {Smethurst}, {Smith}, {Sobeck}, {Spoo}, {Sprague},
  {Stark}, {Stassun}, {Steinmetz}, {Stello}, {Stone-Martinez},
  {Storchi-Bergmann}, {Stringfellow}, {Stutz}, {Su}, {Taghizadeh-Popp},
  {Talbot}, {Tayar}, {Telles}, {Teske}, {Thakar}, {Theissen}, {Tkachenko},
  {Thomas}, {Tojeiro}, {Hernandez Toledo}, {Troup}, {Trump}, {Trussler},
  {Turner}, {Tuttle}, {Unda-Sanzana}, {V{\'a}zquez-Mata}, {Valentini},
  {Valenzuela}, {Vargas-Gonz{\'a}lez}, {Vargas-Maga{\~n}a}, {Alfaro},
  {Villanova}, {Vincenzo}, {Wake}, {Warfield}, {Washington}, {Weaver},
  {Weijmans}, {Weinberg}, {Weiss}, {Westfall}, {Wild}, {Wilde}, {Wilson},
  {Wilson}, {Wilson}, {Wolf}, {Wood-Vasey}, {Yan}, {Zamora}, {Zasowski},
  {Zhang}, {Zhao}, {Zheng}, {Zheng}, \& {Zhu}}]{DR17}
{Abdurro'uf}, {Accetta}, K., {Aerts}, C., {et~al.} 2022, \apjs, 259, 35

\bibitem[{{Andrae} {et~al.}(2023){Andrae}, {Rix}, \& {Chandra}}]{Andrae2023}
{Andrae}, R., {Rix}, H.-W., \& {Chandra}, V. 2023, arXiv e-prints,
  arXiv:2302.02611

\bibitem[{{Andrae} {et~al.}(2022){Andrae}, {Fouesneau}, {Sordo},
  {Bailer-Jones}, {Dharmawardena}, {Rybizki}, {De Angeli}, {Lindstr{\o}m},
  {Marshall}, {Drimmel}, {Korn}, {Soubiran}, {Brouillet}, {Casamiquela}, {Rix},
  {Abreu Aramburu}, {{\'A}lvarez}, {Bakker}, {Bellas-Velidis}, {Bijaoui},
  {Brugaletta}, {Burlacu}, {Carballo}, {Chaoul}, {Chiavassa}, {Contursi},
  {Cooper}, {Creevey}, {Dafonte}, {Dapergolas}, {de Laverny}, {Delchambre},
  {Demouchy}, {Edvardsson}, {Fr{\'e}mat}, {Garabato}, {Garc{\'\i}a-Lario},
  {Garc{\'\i}a-Torres}, {Gavel}, {Gomez}, {Gonz{\'a}lez-Santamar{\'\i}a},
  {Hatzidimitriou}, {Heiter}, {Jean-Antoine Piccolo}, {Kontizas}, {Kordopatis},
  {Lanzafame}, {Lebreton}, {Licata}, {Livanou}, {Lobel}, {Lorca}, {Magdaleno
  Romeo}, {Manteiga}, {Marocco}, {Mary}, {Nicolas}, {Ordenovic}, {Pailler},
  {Palicio}, {Pallas-Quintela}, {Panem}, {Pichon}, {Poggio}, {Recio-Blanco},
  {Riclet}, {Robin}, {Santove{\~n}a}, {Sarro}, {Schultheis}, {Segol},
  {Silvelo}, {Slezak}, {Smart}, {S{\"u}veges}, {Th{\'e}venin}, {Torralba
  Elipe}, {Ulla}, {Utrilla}, {Vallenari}, {van Dillen}, {Zhao}, \&
  {Zorec}}]{Andrae2022}
{Andrae}, R., {Fouesneau}, M., {Sordo}, R., {et~al.} 2022, arXiv e-prints,
  arXiv:2206.06138

\bibitem[{{Astropy Collaboration} {et~al.}(2013){Astropy Collaboration},
  {Robitaille}, {Tollerud}, {Greenfield}, {Droettboom}, {Bray}, {Aldcroft},
  {Davis}, {Ginsburg}, {Price-Whelan}, {Kerzendorf}, {Conley}, {Crighton},
  {Barbary}, {Muna}, {Ferguson}, {Grollier}, {Parikh}, {Nair}, {Unther},
  {Deil}, {Woillez}, {Conseil}, {Kramer}, {Turner}, {Singer}, {Fox}, {Weaver},
  {Zabalza}, {Edwards}, {Azalee Bostroem}, {Burke}, {Casey}, {Crawford},
  {Dencheva}, {Ely}, {Jenness}, {Labrie}, {Lim}, {Pierfederici}, {Pontzen},
  {Ptak}, {Refsdal}, {Servillat}, \& {Streicher}}]{robitaille2013astropy}
{Astropy Collaboration}, {Robitaille}, T.~P., {Tollerud}, E.~J., {et~al.} 2013,
  \aap, 558, A33

\bibitem[{{Astropy Collaboration} {et~al.}(2018){Astropy Collaboration},
  {Price-Whelan}, {Sip{\H{o}}cz}, {G{\"u}nther}, {Lim}, {Crawford}, {Conseil},
  {Shupe}, {Craig}, {Dencheva}, {Ginsburg}, {VanderPlas}, {Bradley},
  {P{\'e}rez-Su{\'a}rez}, {de Val-Borro}, {Aldcroft}, {Cruz}, {Robitaille},
  {Tollerud}, {Ardelean}, {Babej}, {Bach}, {Bachetti}, {Bakanov}, {Bamford},
  {Barentsen}, {Barmby}, {Baumbach}, {Berry}, {Biscani}, {Boquien}, {Bostroem},
  {Bouma}, {Brammer}, {Bray}, {Breytenbach}, {Buddelmeijer}, {Burke},
  {Calderone}, {Cano Rodr{\'\i}guez}, {Cara}, {Cardoso}, {Cheedella}, {Copin},
  {Corrales}, {Crichton}, {D'Avella}, {Deil}, {Depagne}, {Dietrich}, {Donath},
  {Droettboom}, {Earl}, {Erben}, {Fabbro}, {Ferreira}, {Finethy}, {Fox},
  {Garrison}, {Gibbons}, {Goldstein}, {Gommers}, {Greco}, {Greenfield},
  {Groener}, {Grollier}, {Hagen}, {Hirst}, {Homeier}, {Horton}, {Hosseinzadeh},
  {Hu}, {Hunkeler}, {Ivezi{\'c}}, {Jain}, {Jenness}, {Kanarek}, {Kendrew},
  {Kern}, {Kerzendorf}, {Khvalko}, {King}, {Kirkby}, {Kulkarni}, {Kumar},
  {Lee}, {Lenz}, {Littlefair}, {Ma}, {Macleod}, {Mastropietro}, {McCully},
  {Montagnac}, {Morris}, {Mueller}, {Mumford}, {Muna}, {Murphy}, {Nelson},
  {Nguyen}, {Ninan}, {N{\"o}the}, {Ogaz}, {Oh}, {Parejko}, {Parley}, {Pascual},
  {Patil}, {Patil}, {Plunkett}, {Prochaska}, {Rastogi}, {Reddy Janga},
  {Sabater}, {Sakurikar}, {Seifert}, {Sherbert}, {Sherwood-Taylor}, {Shih},
  {Sick}, {Silbiger}, {Singanamalla}, {Singer}, {Sladen}, {Sooley},
  {Sornarajah}, {Streicher}, {Teuben}, {Thomas}, {Tremblay}, {Turner},
  {Terr{\'o}n}, {van Kerkwijk}, {de la Vega}, {Watkins}, {Weaver}, {Whitmore},
  {Woillez}, {Zabalza}, \& {Astropy Contributors}}]{pricewhelan2018astropy}
{Astropy Collaboration}, {Price-Whelan}, A.~M., {Sip{\H{o}}cz}, B.~M., {et~al.}
  2018, \aj, 156, 123

\bibitem[{{Barbosa} {et~al.}(2022){Barbosa}, {Santucci}, {Rossi}, {Limberg},
  {P{\'e}rez-Villegas}, \& {Perottoni}}]{Barbosa2022}
{Barbosa}, F.~O., {Santucci}, R.~M., {Rossi}, S., {et~al.} 2022, \apj, 940, 30

\bibitem[{{Bechtol} {et~al.}(2015){Bechtol}, {Drlica-Wagner}, {Balbinot},
  {Pieres}, {Simon}, {Yanny}, {Santiago}, {Wechsler}, {Frieman}, {Walker},
  {Williams}, {Rozo}, {Rykoff}, {Queiroz}, {Luque}, {Benoit-L{\'e}vy},
  {Tucker}, {Sevilla}, {Gruendl}, {da Costa}, {Fausti Neto}, {Maia}, {Abbott},
  {Allam}, {Armstrong}, {Bauer}, {Bernstein}, {Bernstein}, {Bertin}, {Brooks},
  {Buckley-Geer}, {Burke}, {Carnero Rosell}, {Castander}, {Covarrubias},
  {D'Andrea}, {DePoy}, {Desai}, {Diehl}, {Eifler}, {Estrada}, {Evrard},
  {Fernandez}, {Finley}, {Flaugher}, {Gaztanaga}, {Gerdes}, {Girardi},
  {Gladders}, {Gruen}, {Gutierrez}, {Hao}, {Honscheid}, {Jain}, {James},
  {Kent}, {Kron}, {Kuehn}, {Kuropatkin}, {Lahav}, {Li}, {Lin}, {Makler},
  {March}, {Marshall}, {Martini}, {Merritt}, {Miller}, {Miquel}, {Mohr},
  {Neilsen}, {Nichol}, {Nord}, {Ogando}, {Peoples}, {Petravick}, {Plazas},
  {Romer}, {Roodman}, {Sako}, {Sanchez}, {Scarpine}, {Schubnell}, {Smith},
  {Soares-Santos}, {Sobreira}, {Suchyta}, {Swanson}, {Tarle}, {Thaler},
  {Thomas}, {Wester}, {Zuntz}, \& {DES Collaboration}}]{Bechtol2015}
{Bechtol}, K., {Drlica-Wagner}, A., {Balbinot}, E., {et~al.} 2015, \apj, 807,
  50

\bibitem[{{Belokurov} {et~al.}(2006){Belokurov}, {Zucker}, {Evans}, {Gilmore},
  {Vidrih}, {Bramich}, {Newberg}, {Wyse}, {Irwin}, {Fellhauer}, {Hewett},
  {Walton}, {Wilkinson}, {Cole}, {Yanny}, {Rockosi}, {Beers}, {Bell},
  {Brinkmann}, {Ivezi{\'c}}, \& {Lupton}}]{Belokurov2006}
{Belokurov}, V., {Zucker}, D.~B., {Evans}, N.~W., {et~al.} 2006, \apjl, 642,
  L137

\bibitem[{{Besla} {et~al.}(2012){Besla}, {Kallivayalil}, {Hernquist}, {van der
  Marel}, {Cox}, \& {Kere{\v s}}}]{Besla2012}
{Besla}, G., {Kallivayalil}, N., {Hernquist}, L., {et~al.} 2012, \mnras, 421,
  2109

\bibitem[{{Chambers} {et~al.}(2016){Chambers}, {Magnier}, {Metcalfe},
  {Flewelling}, {Huber}, {Waters}, {Denneau}, {Draper}, {Farrow}, {Finkbeiner},
  {Holmberg}, {Koppenhoefer}, {Price}, {Rest}, {Saglia}, {Schlafly}, {Smartt},
  {Sweeney}, {Wainscoat}, {Burgett}, {Chastel}, {Grav}, {Heasley}, {Hodapp},
  {Jedicke}, {Kaiser}, {Kudritzki}, {Luppino}, {Lupton}, {Monet}, {Morgan},
  {Onaka}, {Shiao}, {Stubbs}, {Tonry}, {White}, {Ba{\~n}ados}, {Bell},
  {Bender}, {Bernard}, {Boegner}, {Boffi}, {Botticella}, {Calamida},
  {Casertano}, {Chen}, {Chen}, {Cole}, {Deacon}, {Frenk}, {Fitzsimmons},
  {Gezari}, {Gibbs}, {Goessl}, {Goggia}, {Gourgue}, {Goldman}, {Grant},
  {Grebel}, {Hambly}, {Hasinger}, {Heavens}, {Heckman}, {Henderson}, {Henning},
  {Holman}, {Hopp}, {Ip}, {Isani}, {Jackson}, {Keyes}, {Koekemoer}, {Kotak},
  {Le}, {Liska}, {Long}, {Lucey}, {Liu}, {Martin}, {Masci}, {McLean}, {Mindel},
  {Misra}, {Morganson}, {Murphy}, {Obaika}, {Narayan}, {Nieto-Santisteban},
  {Norberg}, {Peacock}, {Pier}, {Postman}, {Primak}, {Rae}, {Rai}, {Riess},
  {Riffeser}, {Rix}, {R{\"o}ser}, {Russel}, {Rutz}, {Schilbach}, {Schultz},
  {Scolnic}, {Strolger}, {Szalay}, {Seitz}, {Small}, {Smith}, {Soderblom},
  {Taylor}, {Thomson}, {Taylor}, {Thakar}, {Thiel}, {Thilker}, {Unger},
  {Urata}, {Valenti}, {Wagner}, {Walder}, {Walter}, {Watters}, {Werner},
  {Wood-Vasey}, \& {Wyse}}]{Chambers2016}
{Chambers}, K.~C., {Magnier}, E.~A., {Metcalfe}, N., {et~al.} 2016, arXiv
  e-prints, arXiv:1612.05560

\bibitem[{Chen \& Guestrin(2016)}]{Chen2016}
Chen, T., \& Guestrin, C. 2016, in Proceedings of the 22nd ACM SIGKDD
  International Conference on Knowledge Discovery and Data Mining, KDD '16 (New
  York, NY, USA: ACM), 785--794.
\newblock \url{http://doi.acm.org/10.1145/2939672.2939785}

\bibitem[{{Conn} {et~al.}(2008){Conn}, {Lane}, {Lewis}, {Irwin}, {Ibata},
  {Martin}, {Bellazzini}, \& {Tuntsov}}]{Conn2008}
{Conn}, B.~C., {Lane}, R.~R., {Lewis}, G.~F., {et~al.} 2008, \mnras, 390, 1388

\bibitem[{{Cutri} {et~al.}(2021){Cutri}, {Wright}, {Conrow}, {Fowler},
  {Eisenhardt}, {Grillmair}, {Kirkpatrick}, {Masci}, {McCallon}, {Wheelock},
  {Fajardo-Acosta}, {Yan}, {Benford}, {Harbut}, {Jarrett}, {Lake}, {Leisawitz},
  {Ressler}, {Stanford}, {Tsai}, {Liu}, {Helou}, {Mainzer}, {Gettngs},
  {Gonzalez}, {Hoffman}, {Marsh}, {Padgett}, {Skrutskie}, {Beck}, {Papin}, \&
  {Wittman}}]{Cutri2014}
{Cutri}, R.~M., {Wright}, E.~L., {Conrow}, T., {et~al.} 2021, VizieR Online
  Data Catalog, II/328

\bibitem[{{Dark Energy Survey Collaboration} {et~al.}(2016){Dark Energy Survey
  Collaboration}, {Abbott}, {Abdalla}, {Aleksi{\'c}}, {Allam}, {Amara},
  {Bacon}, {Balbinot}, {Banerji}, {Bechtol}, {Benoit-L{\'e}vy}, {Bernstein},
  {Bertin}, {Blazek}, {Bonnett}, {Bridle}, {Brooks}, {Brunner}, {Buckley-Geer},
  {Burke}, {Caminha}, {Capozzi}, {Carlsen}, {Carnero-Rosell}, {Carollo},
  {Carrasco-Kind}, {Carretero}, {Castander}, {Clerkin}, {Collett}, {Conselice},
  {Crocce}, {Cunha}, {D'Andrea}, {da Costa}, {Davis}, {Desai}, {Diehl},
  {Dietrich}, {Dodelson}, {Doel}, {Drlica-Wagner}, {Estrada}, {Etherington},
  {Evrard}, {Fabbri}, {Finley}, {Flaugher}, {Foley}, {Fosalba}, {Frieman},
  {Garc{\'{\i}}a-Bellido}, {Gaztanaga}, {Gerdes}, {Giannantonio}, {Goldstein},
  {Gruen}, {Gruendl}, {Guarnieri}, {Gutierrez}, {Hartley}, {Honscheid}, {Jain},
  {James}, {Jeltema}, {Jouvel}, {Kessler}, {King}, {Kirk}, {Kron}, {Kuehn},
  {Kuropatkin}, {Lahav}, {Li}, {Lima}, {Lin}, {Maia}, {Makler}, {Manera},
  {Maraston}, {Marshall}, {Martini}, {McMahon}, {Melchior}, {Merson}, {Miller},
  {Miquel}, {Mohr}, {Morice-Atkinson}, {Naidoo}, {Neilsen}, {Nichol}, {Nord},
  {Ogando}, {Ostrovski}, {Palmese}, {Papadopoulos}, {Peiris}, {Peoples},
  {Percival}, {Plazas}, {Reed}, {Refregier}, {Romer}, {Roodman}, {Ross},
  {Rozo}, {Rykoff}, {Sadeh}, {Sako}, {S{\'a}nchez}, {Sanchez}, {Santiago},
  {Scarpine}, {Schubnell}, {Sevilla-Noarbe}, {Sheldon}, {Smith}, {Smith},
  {Soares-Santos}, {Sobreira}, {Soumagnac}, {Suchyta}, {Sullivan}, {Swanson},
  {Tarle}, {Thaler}, {Thomas}, {Thomas}, {Tucker}, {Vieira}, {Vikram},
  {Walker}, {Wechsler}, {Weller}, {Wester}, {Whiteway}, {Wilcox}, {Yanny},
  {Zhang}, \& {Zuntz}}]{DES}
{Dark Energy Survey Collaboration}, {Abbott}, T., {Abdalla}, F.~B., {et~al.}
  2016, \mnras, 460, 1270

\bibitem[{{Drlica-Wagner} {et~al.}(2015){Drlica-Wagner}, {Bechtol}, {Rykoff},
  {Luque}, {Queiroz}, {Mao}, {Wechsler}, {Simon}, {Santiago}, {Yanny},
  {Balbinot}, {Dodelson}, {Fausti Neto}, {James}, {Li}, {Maia}, {Marshall},
  {Pieres}, {Stringer}, {Walker}, {Abbott}, {Abdalla}, {Allam},
  {Benoit-L{\'e}vy}, {Bernstein}, {Bertin}, {Brooks}, {Buckley-Geer}, {Burke},
  {Carnero Rosell}, {Carrasco Kind}, {Carretero}, {Crocce}, {da Costa},
  {Desai}, {Diehl}, {Dietrich}, {Doel}, {Eifler}, {Evrard}, {Finley},
  {Flaugher}, {Fosalba}, {Frieman}, {Gaztanaga}, {Gerdes}, {Gruen}, {Gruendl},
  {Gutierrez}, {Honscheid}, {Kuehn}, {Kuropatkin}, {Lahav}, {Martini},
  {Miquel}, {Nord}, {Ogando}, {Plazas}, {Reil}, {Roodman}, {Sako}, {Sanchez},
  {Scarpine}, {Schubnell}, {Sevilla-Noarbe}, {Smith}, {Soares-Santos},
  {Sobreira}, {Suchyta}, {Swanson}, {Tarle}, {Tucker}, {Vikram}, {Wester},
  {Zhang}, {Zuntz}, \& {DES Collaboration}}]{Drlica-Wagner2015}
{Drlica-Wagner}, A., {Bechtol}, K., {Rykoff}, E.~S., {et~al.} 2015, \apj, 813,
  109

\bibitem[{{Drlica-Wagner} {et~al.}(2021){Drlica-Wagner}, {Carlin}, {Nidever},
  {Ferguson}, {Kuropatkin}, {Adam{\'o}w}, {Cerny}, {Choi}, {Esteves},
  {Mart{\'\i}nez-V{\'a}zquez}, {Mau}, {Miller}, {Mutlu-Pakdil}, {Neilsen},
  {Olsen}, {Pace}, {Riley}, {Sakowska}, {Sand}, {Santana-Silva}, {Tollerud},
  {Tucker}, {Vivas}, {Zaborowski}, {Zenteno}, {Abbott}, {Allam}, {Bechtol},
  {Bell}, {Bell}, {Bilaji}, {Bom}, {Carballo-Bello}, {Crnojevi{\'c}}, {Cioni},
  {Diaz-Ocampo}, {de Boer}, {Erkal}, {Gruendl}, {Hernandez-Lang}, {Hughes},
  {James}, {Johnson}, {Li}, {Mao}, {Mart{\'\i}nez-Delgado}, {Massana},
  {McNanna}, {Morgan}, {Nadler}, {No{\"e}l}, {Palmese}, {Peter}, {Rykoff},
  {S{\'a}nchez}, {Shipp}, {Simon}, {Smercina}, {Soares-Santos}, {Stringfellow},
  {Tavangar}, {van der Marel}, {Walker}, {Wechsler}, {Wu}, {Yanny},
  {Fitzpatrick}, {Huang}, {Jacques}, {Nikutta}, {Scott}, \& {Astro Data
  Lab}}]{DrlicaWagner2021}
{Drlica-Wagner}, A., {Carlin}, J.~L., {Nidever}, D.~L., {et~al.} 2021, \apjs,
  256, 2

\bibitem[{{Erkal} {et~al.}(2019){Erkal}, {Belokurov}, {Laporte}, {Koposov},
  {Li}, {Grillmair}, {Kallivayalil}, {Price-Whelan}, {Evans}, {Hawkins},
  {Hendel}, {Mateu}, {Navarro}, {del Pino}, {Slater}, {Sohn}, \& {Orphan Aspen
  Treasury Collaboration}}]{Erkal2019}
{Erkal}, D., {Belokurov}, V., {Laporte}, C.~F.~P., {et~al.} 2019, \mnras, 487,
  2685

\bibitem[{{Ferguson} {et~al.}(2022){Ferguson}, {Shipp}, {Drlica-Wagner}, {Li},
  {Cerny}, {Tavangar}, {Pace}, {Marshall}, {Riley}, {Adam{\'o}w}, {Carlin},
  {Choi}, {Erkal}, {James}, {Koposov}, {Kuropatkin},
  {Mart{\'\i}nez-V{\'a}zquez}, {Mau}, {Mutlu-Pakdil}, {Olsen}, {Sakowska},
  {Stringfellow}, {Yanny}, \& {Yanny}}]{Ferguson2022}
{Ferguson}, P.~S., {Shipp}, N., {Drlica-Wagner}, A., {et~al.} 2022, \aj, 163,
  18

\bibitem[{{Flaugher} {et~al.}(2015){Flaugher}, {Diehl}, {Honscheid}, {Abbott},
  {Alvarez}, {Angstadt}, {Annis}, {Antonik}, {Ballester}, {Beaufore},
  {Bernstein}, {Bernstein}, {Bigelow}, {Bonati}, {Boprie}, {Brooks},
  {Buckley-Geer}, {Campa}, {Cardiel-Sas}, {Castander}, {Castilla}, {Cease},
  {Cela-Ruiz}, {Chappa}, {Chi}, {Cooper}, {da Costa}, {Dede}, {Derylo},
  {DePoy}, {de Vicente}, {Doel}, {Drlica-Wagner}, {Eiting}, {Elliott}, {Emes},
  {Estrada}, {Fausti Neto}, {Finley}, {Flores}, {Frieman}, {Gerdes},
  {Gladders}, {Gregory}, {Gutierrez}, {Hao}, {Holland}, {Holm}, {Huffman},
  {Jackson}, {James}, {Jonas}, {Karcher}, {Karliner}, {Kent}, {Kessler},
  {Kozlovsky}, {Kron}, {Kubik}, {Kuehn}, {Kuhlmann}, {Kuk}, {Lahav}, {Lathrop},
  {Lee}, {Levi}, {Lewis}, {Li}, {Mandrichenko}, {Marshall}, {Martinez},
  {Merritt}, {Miquel}, {Mu{\~n}oz}, {Neilsen}, {Nichol}, {Nord}, {Ogando},
  {Olsen}, {Palaio}, {Patton}, {Peoples}, {Plazas}, {Rauch}, {Reil}, {Rheault},
  {Roe}, {Rogers}, {Roodman}, {Sanchez}, {Scarpine}, {Schindler}, {Schmidt},
  {Schmitt}, {Schubnell}, {Schultz}, {Schurter}, {Scott}, {Serrano}, {Shaw},
  {Smith}, {Soares-Santos}, {Stefanik}, {Stuermer}, {Suchyta}, {Sypniewski},
  {Tarle}, {Thaler}, {Tighe}, {Tran}, {Tucker}, {Walker}, {Wang}, {Watson},
  {Weaverdyck}, {Wester}, {Woods}, {Yanny}, \& {DES
  Collaboration}}]{Flaugher2015}
{Flaugher}, B., {Diehl}, H.~T., {Honscheid}, K., {et~al.} 2015, \aj, 150, 150

\bibitem[{{Gaia Collaboration} {et~al.}(2018){Gaia Collaboration}, {Brown},
  {Vallenari}, {Prusti}, {de Bruijne}, {Babusiaux}, {Bailer-Jones}, {Biermann},
  {Evans}, {Eyer}, {Jansen}, {Jordi}, {Klioner}, {Lammers}, {Lindegren},
  {Luri}, {Mignard}, {Panem}, {Pourbaix}, {Randich}, {Sartoretti}, {Siddiqui},
  {Soubiran}, {van Leeuwen}, {Walton}, {Arenou}, {Bastian}, {Cropper},
  {Drimmel}, {Katz}, {Lattanzi}, {Bakker}, {Cacciari}, {Casta{\~n}eda},
  {Chaoul}, {Cheek}, {De Angeli}, {Fabricius}, {Guerra}, {Holl}, {Masana},
  {Messineo}, {Mowlavi}, {Nienartowicz}, {Panuzzo}, {Portell}, {Riello},
  {Seabroke}, {Tanga}, {Th{\'e}venin}, {Gracia-Abril}, {Comoretto},
  {Garcia-Reinaldos}, {Teyssier}, {Altmann}, {Andrae}, {Audard},
  {Bellas-Velidis}, {Benson}, {Berthier}, {Blomme}, {Burgess}, {Busso},
  {Carry}, {Cellino}, {Clementini}, {Clotet}, {Creevey}, {Davidson}, {De
  Ridder}, {Delchambre}, {Dell'Oro}, {Ducourant},
  {Fern{\'a}ndez-Hern{\'a}ndez}, {Fouesneau}, {Fr{\'e}mat}, {Galluccio},
  {Garc{\'\i}a-Torres}, {Gonz{\'a}lez-N{\'u}{\~n}ez}, {Gonz{\'a}lez-Vidal},
  {Gosset}, {Guy}, {Halbwachs}, {Hambly}, {Harrison}, {Hern{\'a}ndez},
  {Hestroffer}, {Hodgkin}, {Hutton}, {Jasniewicz}, {Jean-Antoine-Piccolo},
  {Jordan}, {Korn}, {Krone-Martins}, {Lanzafame}, {Lebzelter}, {L{\"o}ffler},
  {Manteiga}, {Marrese}, {Mart{\'\i}n-Fleitas}, {Moitinho}, {Mora}, {Muinonen},
  {Osinde}, {Pancino}, {Pauwels}, {Petit}, {Recio-Blanco}, {Richards},
  {Rimoldini}, {Robin}, {Sarro}, {Siopis}, {Smith}, {Sozzetti}, {S{\"u}veges},
  {Torra}, {van Reeven}, {Abbas}, {Abreu Aramburu}, {Accart}, {Aerts},
  {Altavilla}, {{\'A}lvarez}, {Alvarez}, {Alves}, {Anderson}, {Andrei},
  {Anglada Varela}, {Antiche}, {Antoja}, {Arcay}, {Astraatmadja}, {Bach},
  {Baker}, {Balaguer-N{\'u}{\~n}ez}, {Balm}, {Barache}, {Barata}, {Barbato},
  {Barblan}, {Barklem}, {Barrado}, {Barros}, {Barstow}, {Bartholom{\'e}
  Mu{\~n}oz}, {Bassilana}, {Becciani}, {Bellazzini}, {Berihuete}, {Bertone},
  {Bianchi}, {Bienaym{\'e}}, {Blanco-Cuaresma}, {Boch}, {Boeche}, {Bombrun},
  {Borrachero}, {Bossini}, {Bouquillon}, {Bourda}, {Bragaglia}, {Bramante},
  {Breddels}, {Bressan}, {Brouillet}, {Br{\"u}semeister}, {Brugaletta},
  {Bucciarelli}, {Burlacu}, {Busonero}, {Butkevich}, {Buzzi}, {Caffau},
  {Cancelliere}, {Cannizzaro}, {Cantat-Gaudin}, {Carballo}, {Carlucci},
  {Carrasco}, {Casamiquela}, {Castellani}, {Castro-Ginard}, {Charlot},
  {Chemin}, {Chiavassa}, {Cocozza}, {Costigan}, {Cowell}, {Crifo}, {Crosta},
  {Crowley}, {Cuypers}, {Dafonte}, {Damerdji}, {Dapergolas}, {David}, {David},
  {de Laverny}, {De Luise}, {De March}, {de Martino}, {de Souza}, {de Torres},
  {Debosscher}, {del Pozo}, {Delbo}, {Delgado}, {Delgado}, {Di Matteo},
  {Diakite}, {Diener}, {Distefano}, {Dolding}, {Drazinos}, {Dur{\'a}n},
  {Edvardsson}, {Enke}, {Eriksson}, {Esquej}, {Eynard Bontemps}, {Fabre},
  {Fabrizio}, {Faigler}, {Falc{\~a}o}, {Farr{\`a}s Casas}, {Federici},
  {Fedorets}, {Fernique}, {Figueras}, {Filippi}, {Findeisen}, {Fonti},
  {Fraile}, {Fraser}, {Fr{\'e}zouls}, {Gai}, {Galleti}, {Garabato},
  {Garc{\'\i}a-Sedano}, {Garofalo}, {Garralda}, {Gavel}, {Gavras}, {Gerssen},
  {Geyer}, {Giacobbe}, {Gilmore}, {Girona}, {Giuffrida}, {Glass}, {Gomes},
  {Granvik}, {Gueguen}, {Guerrier}, {Guiraud}, {Guti{\'e}rrez-S{\'a}nchez},
  {Haigron}, {Hatzidimitriou}, {Hauser}, {Haywood}, {Heiter}, {Helmi}, {Heu},
  {Hilger}, {Hobbs}, {Hofmann}, {Holland}, {Huckle}, {Hypki}, {Icardi},
  {Jan{\ss}en}, {Jevardat de Fombelle}, {Jonker}, {Juh{\'a}sz}, {Julbe},
  {Karampelas}, {Kewley}, {Klar}, {Kochoska}, {Kohley}, {Kolenberg},
  {Kontizas}, {Kontizas}, {Koposov}, {Kordopatis}, {Kostrzewa-Rutkowska},
  {Koubsky}, {Lambert}, {Lanza}, {Lasne}, {Lavigne}, {Le Fustec}, {Le
  Poncin-Lafitte}, {Lebreton}, {Leccia}, {Leclerc}, {Lecoeur-Taibi},
  {Lenhardt}, {Leroux}, {Liao}, {Licata}, {Lindstr{\o}m}, {Lister}, {Livanou},
  {Lobel}, {L{\'o}pez}, {Managau}, {Mann}, {Mantelet}, {Marchal}, {Marchant},
  {Marconi}, {Marinoni}, {Marschalk{\'o}}, {Marshall}, {Martino}, {Marton},
  {Mary}, {Massari}, {Matijevi{\v{c}}}, {Mazeh}, {McMillan}, {Messina},
  {Michalik}, {Millar}, {Molina}, {Molinaro}, {Moln{\'a}r}, {Montegriffo},
  {Mor}, {Morbidelli}, {Morel}, {Morris}, {Mulone}, {Muraveva}, {Musella},
  {Nelemans}, {Nicastro}, {Noval}, {O'Mullane}, {Ord{\'e}novic},
  {Ord{\'o}{\~n}ez-Blanco}, {Osborne}, {Pagani}, {Pagano}, {Pailler},
  {Palacin}, {Palaversa}, {Panahi}, {Pawlak}, {Piersimoni}, {Pineau}, {Plachy},
  {Plum}, {Poggio}, {Poujoulet}, {Pr{\v{s}}a}, {Pulone}, {Racero}, {Ragaini},
  {Rambaux}, {Ramos-Lerate}, {Regibo}, {Reyl{\'e}}, {Riclet}, {Ripepi}, {Riva},
  {Rivard}, {Rixon}, {Roegiers}, {Roelens}, {Romero-G{\'o}mez}, {Rowell},
  {Royer}, {Ruiz-Dern}, {Sadowski}, {Sagrist{\`a} Sell{\'e}s}, {Sahlmann},
  {Salgado}, {Salguero}, {Sanna}, {Santana-Ros}, {Sarasso}, {Savietto},
  {Schultheis}, {Sciacca}, {Segol}, {Segovia}, {S{\'e}gransan}, {Shih},
  {Siltala}, {Silva}, {Smart}, {Smith}, {Solano}, {Solitro}, {Sordo}, {Soria
  Nieto}, {Souchay}, {Spagna}, {Spoto}, {Stampa}, {Steele},
  {Steidelm{\"u}ller}, {Stephenson}, {Stoev}, {Suess}, {Surdej}, {Szabados},
  {Szegedi-Elek}, {Tapiador}, {Taris}, {Tauran}, {Taylor}, {Teixeira},
  {Terrett}, {Teyssandier}, {Thuillot}, {Titarenko}, {Torra Clotet}, {Turon},
  {Ulla}, {Utrilla}, {Uzzi}, {Vaillant}, {Valentini}, {Valette}, {van Elteren},
  {Van Hemelryck}, {van Leeuwen}, {Vaschetto}, {Vecchiato}, {Veljanoski},
  {Viala}, {Vicente}, {Vogt}, {von Essen}, {Voss}, {Votruba}, {Voutsinas},
  {Walmsley}, {Weiler}, {Wertz}, {Wevers}, {Wyrzykowski}, {Yoldas},
  {{\v{Z}}erjal}, {Ziaeepour}, {Zorec}, {Zschocke}, {Zucker}, {Zurbach}, \&
  {Zwitter}}]{Brown2018}
{Gaia Collaboration}, {Brown}, A.~G.~A., {Vallenari}, A., {et~al.} 2018, \aap,
  616, A1

\bibitem[{{Gaia Collaboration} {et~al.}(2022){Gaia Collaboration}, {Vallenari},
  {Brown}, {Prusti}, {de Bruijne}, {Arenou}, {Babusiaux}, {Biermann},
  {Creevey}, {Ducourant}, {Evans}, {Eyer}, {Guerra}, {Hutton}, {Jordi},
  {Klioner}, {Lammers}, {Lindegren}, {Luri}, {Mignard}, {Panem}, {Pourbaix},
  {Randich}, {Sartoretti}, {Soubiran}, {Tanga}, {Walton}, {Bailer-Jones},
  {Bastian}, {Drimmel}, {Jansen}, {Katz}, {Lattanzi}, {van Leeuwen}, {Bakker},
  {Cacciari}, {Casta{\~n}eda}, {De Angeli}, {Fabricius}, {Fouesneau},
  {Fr{\'e}mat}, {Galluccio}, {Guerrier}, {Heiter}, {Masana}, {Messineo},
  {Mowlavi}, {Nicolas}, {Nienartowicz}, {Pailler}, {Panuzzo}, {Riclet}, {Roux},
  {Seabroke}, {Sordo{\o}rcit}, {Th{\'e}venin}, {Gracia-Abril}, {Portell},
  {Teyssier}, {Altmann}, {Andrae}, {Audard}, {Bellas-Velidis}, {Benson},
  {Berthier}, {Blomme}, {Burgess}, {Busonero}, {Busso}, {C{\'a}novas}, {Carry},
  {Cellino}, {Cheek}, {Clementini}, {Damerdji}, {Davidson}, {de Teodoro},
  {Nu{\~n}ez Campos}, {Delchambre}, {Dell'Oro}, {Esquej},
  {Fern{\'a}ndez-Hern{\'a}ndez}, {Fraile}, {Garabato}, {Garc{\'\i}a-Lario},
  {Gosset}, {Haigron}, {Halbwachs}, {Hambly}, {Harrison}, {Hern{\'a}ndez},
  {Hestroffer}, {Hodgkin}, {Holl}, {Jan{\ss}en}, {Jevardat de Fombelle},
  {Jordan}, {Krone-Martins}, {Lanzafame}, {L{\"o}ffler}, {Marchal}, {Marrese},
  {Moitinho}, {Muinonen}, {Osborne}, {Pancino}, {Pauwels}, {Recio-Blanco},
  {Reyl{\'e}}, {Riello}, {Rimoldini}, {Roegiers}, {Rybizki}, {Sarro}, {Siopis},
  {Smith}, {Sozzetti}, {Utrilla}, {van Leeuwen}, {Abbas}, {{\'A}brah{\'a}m},
  {Abreu Aramburu}, {Aerts}, {Aguado}, {Ajaj}, {Aldea-Montero}, {Altavilla},
  {{\'A}lvarez}, {Alves}, {Anders}, {Anderson}, {Anglada Varela}, {Antoja},
  {Baines}, {Baker}, {Balaguer-N{\'u}{\~n}ez}, {Balbinot}, {Balog}, {Barache},
  {Barbato}, {Barros}, {Barstow}, {Bartolom{\'e}}, {Bassilana}, {Bauchet},
  {Becciani}, {Bellazzini}, {Berihuete}, {Bernet}, {Bertone}, {Bianchi},
  {Binnenfeld}, {Blanco-Cuaresma}, {Blazere}, {Boch}, {Bombrun}, {Bossini},
  {Bouquillon}, {Bragaglia}, {Bramante}, {Breedt}, {Bressan}, {Brouillet},
  {Brugaletta}, {Bucciarelli}, {Burlacu}, {Butkevich}, {Buzzi}, {Caffau},
  {Cancelliere}, {Cantat-Gaudin}, {Carballo}, {Carlucci}, {Carnerero},
  {Carrasco}, {Casamiquela}, {Castellani}, {Castro-Ginard}, {Chaoul},
  {Charlot}, {Chemin}, {Chiaramida}, {Chiavassa}, {Chornay}, {Comoretto},
  {Contursi}, {Cooper}, {Cornez}, {Cowell}, {Crifo}, {Cropper}, {Crosta},
  {Crowley}, {Dafonte}, {Dapergolas}, {David}, {David}, {de Laverny}, {De
  Luise}, {De March}, {De Ridder}, {de Souza}, {de Torres}, {del Peloso}, {del
  Pozo}, {Delbo}, {Delgado}, {Delisle}, {Demouchy}, {Dharmawardena}, {Di
  Matteo}, {Diakite}, {Diener}, {Distefano}, {Dolding}, {Edvardsson}, {Enke},
  {Fabre}, {Fabrizio}, {Faigler}, {Fedorets}, {Fernique}, {Fienga}, {Figueras},
  {Fournier}, {Fouron}, {Fragkoudi}, {Gai}, {Garcia-Gutierrez},
  {Garcia-Reinaldos}, {Garc{\'\i}a-Torres}, {Garofalo}, {Gavel}, {Gavras},
  {Gerlach}, {Geyer}, {Giacobbe}, {Gilmore}, {Girona}, {Giuffrida}, {Gomel},
  {Gomez}, {Gonz{\'a}lez-N{\'u}{\~n}ez}, {Gonz{\'a}lez-Santamar{\'\i}a},
  {Gonz{\'a}lez-Vidal}, {Granvik}, {Guillout}, {Guiraud},
  {Guti{\'e}rrez-S{\'a}nchez}, {Guy}, {Hatzidimitriou}, {Hauser}, {Haywood},
  {Helmer}, {Helmi}, {Sarmiento}, {Hidalgo}, {Hilger}, {H{\l}adczuk}, {Hobbs},
  {Holland}, {Huckle}, {Jardine}, {Jasniewicz}, {Jean-Antoine Piccolo},
  {Jim{\'e}nez-Arranz}, {Jorissen}, {Juaristi Campillo}, {Julbe}, {Karbevska},
  {Kervella}, {Khanna}, {Kontizas}, {Kordopatis}, {Korn}, {K{\'o}sp{\'a}l},
  {Kostrzewa-Rutkowska}, {Kruszy{\'n}ska}, {Kun}, {Laizeau}, {Lambert},
  {Lanza}, {Lasne}, {Le Campion}, {Lebreton}, {Lebzelter}, {Leccia}, {Leclerc},
  {Lecoeur-Taibi}, {Liao}, {Licata}, {Lindstr{\o}m}, {Lister}, {Livanou},
  {Lobel}, {Lorca}, {Loup}, {Madrero Pardo}, {Magdaleno Romeo}, {Managau},
  {Mann}, {Manteiga}, {Marchant}, {Marconi}, {Marcos}, {Marcos Santos},
  {Mar{\'\i}n Pina}, {Marinoni}, {Marocco}, {Marshall}, {Polo},
  {Mart{\'\i}n-Fleitas}, {Marton}, {Mary}, {Masip}, {Massari},
  {Mastrobuono-Battisti}, {Mazeh}, {McMillan}, {Messina}, {Michalik}, {Millar},
  {Mints}, {Molina}, {Molinaro}, {Moln{\'a}r}, {Monari}, {Mongui{\'o}},
  {Montegriffo}, {Montero}, {Mor}, {Mora}, {Morbidelli}, {Morel}, {Morris},
  {Muraveva}, {Murphy}, {Musella}, {Nagy}, {Noval}, {Oca{\~n}a}, {Ogden},
  {Ordenovic}, {Osinde}, {Pagani}, {Pagano}, {Palaversa}, {Palicio},
  {Pallas-Quintela}, {Panahi}, {Payne-Wardenaar}, {Pe{\~n}alosa Esteller},
  {Penttil{\"a}}, {Pichon}, {Piersimoni}, {Pineau}, {Plachy}, {Plum}, {Poggio},
  {Pr{\v{s}}a}, {Pulone}, {Racero}, {Ragaini}, {Rainer}, {Raiteri}, {Rambaux},
  {Ramos}, {Ramos-Lerate}, {Re Fiorentin}, {Regibo}, {Richards}, {Rios Diaz},
  {Ripepi}, {Riva}, {Rix}, {Rixon}, {Robichon}, {Robin}, {Robin}, {Roelens},
  {Rogues}, {Rohrbasser}, {Romero-G{\'o}mez}, {Rowell}, {Royer}, {Ruz Mieres},
  {Rybicki}, {Sadowski}, {S{\'a}ez N{\'u}{\~n}ez}, {Sagrist{\`a} Sell{\'e}s},
  {Sahlmann}, {Salguero}, {Samaras}, {Sanchez Gimenez}, {Sanna},
  {Santove{\~n}a}, {Sarasso}, {Schultheis}, {Sciacca}, {Segol}, {Segovia},
  {S{\'e}gransan}, {Semeux}, {Shahaf}, {Siddiqui}, {Siebert}, {Siltala},
  {Silvelo}, {Slezak}, {Slezak}, {Smart}, {Snaith}, {Solano}, {Solitro},
  {Souami}, {Souchay}, {Spagna}, {Spina}, {Spoto}, {Steele},
  {Steidelm{\"u}ller}, {Stephenson}, {S{\"u}veges}, {Surdej}, {Szabados},
  {Szegedi-Elek}, {Taris}, {Taylo}, {Teixeira}, {Tolomei}, {Tonello}, {Torra},
  {Torra}, {Torralba Elipe}, {Trabucchi}, {Tsounis}, {Turon}, {Ulla}, {Unger},
  {Vaillant}, {van Dillen}, {van Reeven}, {Vanel}, {Vecchiato}, {Viala},
  {Vicente}, {Voutsinas}, {Weiler}, {Wevers}, {Wyrzykowski}, {Yoldas}, {Yvard},
  {Zhao}, {Zorec}, {Zucker}, \& {Zwitter}}]{Vallenari2022}
{Gaia Collaboration}, {Vallenari}, A., {Brown}, A.~G.~A., {et~al.} 2022, arXiv
  e-prints, arXiv:2208.00211

\bibitem[{{Grillmair}(2006)}]{Grillmair2006e}
{Grillmair}, C.~J. 2006, \apjl, 651, L29

\bibitem[{{Grillmair} \& {Dionatos}(2006{\natexlab{a}})}]{Grillmair2006a}
{Grillmair}, C.~J., \& {Dionatos}, O. 2006{\natexlab{a}}, \apjl, 641, L37

\bibitem[{{Grillmair} \& {Dionatos}(2006{\natexlab{b}})}]{Grillmair2006b}
---. 2006{\natexlab{b}}, \apjl, 643, L17

\bibitem[{Harris {et~al.}(2020)Harris, Millman, van~der Walt, Gommers,
  Virtanen, Cournapeau, Wieser, Taylor, Berg, Smith, Kern, Picus, Hoyer, van
  Kerkwijk, Brett, Haldane, del R{\'{i}}o, Wiebe, Peterson,
  G{\'{e}}rard-Marchant, Sheppard, Reddy, Weckesser, Abbasi, Gohlke, \&
  Oliphant}]{harris2020numpy}
Harris, C.~R., Millman, K.~J., van~der Walt, S.~J., {et~al.} 2020, Nature, 585,
  357.
\newblock \url{https://doi.org/10.1038/s41586-020-2649-2}

\bibitem[{{Hayes} {et~al.}(2018){Hayes}, {Majewski}, {Hasselquist}, {Beaton},
  {Cunha}, {Smith}, {Price-Whelan}, {Anguiano}, {Beers}, {Carrera},
  {Fern{\'a}ndez-Trincado}, {Frinchaboy}, {Garc{\'{\i}}a-Hern{\'a}ndez},
  {Lane}, {Nidever}, {Nitschelm}, {Roman-Lopes}, \& {Zamora}}]{Hayes2018b}
{Hayes}, C.~R., {Majewski}, S.~R., {Hasselquist}, S., {et~al.} 2018, \apjl,
  859, L8

\bibitem[{Hunter(2007)}]{hunter2007matplotlib}
Hunter, J.~D. 2007, Computing in Science \& Engineering, 9, 90

\bibitem[{{Ibata} {et~al.}(2001){Ibata}, {Irwin}, {Lewis}, \&
  {Stolte}}]{Ibata2001}
{Ibata}, R., {Irwin}, M., {Lewis}, G.~F., \& {Stolte}, A. 2001, \apjl, 547,
  L133

\bibitem[{{Ibata} {et~al.}(2019){Ibata}, {Malhan}, \& {Martin}}]{Ibata2019}
{Ibata}, R.~A., {Malhan}, K., \& {Martin}, N.~F. 2019, \apj, 872, 152

\bibitem[{{Johnston} {et~al.}(2005){Johnston}, {Law}, \&
  {Majewski}}]{Johnston2005}
{Johnston}, K.~V., {Law}, D.~R., \& {Majewski}, S.~R. 2005, \apj, 619, 800

\bibitem[{{Koposov} {et~al.}(2010){Koposov}, {Rix}, \& {Hogg}}]{Koposov2010}
{Koposov}, S.~E., {Rix}, H.-W., \& {Hogg}, D.~W. 2010, \apj, 712, 260

\bibitem[{{Koposov} {et~al.}(2012){Koposov}, {Belokurov}, {Evans}, {Gilmore},
  {Gieles}, {Irwin}, {Lewis}, {Niederste-Ostholt}, {Pe{\~n}arrubia}, {Smith},
  {Bizyaev}, {Malanushenko}, {Malanushenko}, {Schneider}, \&
  {Wyse}}]{Koposov2012}
{Koposov}, S.~E., {Belokurov}, V., {Evans}, N.~W., {et~al.} 2012, \apj, 750, 80

\bibitem[{{Koposov} {et~al.}(2018){Koposov}, {Walker}, {Belokurov}, {Casey},
  {Geringer-Sameth}, {Mackey}, {Da Costa}, {Erkal}, {Jethwa}, {Mateo},
  {Olszewski}, \& {Bailey}}]{Koposov2018}
{Koposov}, S.~E., {Walker}, M.~G., {Belokurov}, V., {et~al.} 2018, \mnras, 479,
  5343

\bibitem[{{Majewski} {et~al.}(2003){Majewski}, {Skrutskie}, {Weinberg}, \&
  {Ostheimer}}]{Majewski2003}
{Majewski}, S.~R., {Skrutskie}, M.~F., {Weinberg}, M.~D., \& {Ostheimer}, J.~C.
  2003, \apj, 599, 1082

\bibitem[{{Majewski} {et~al.}(2017){Majewski}, {Schiavon}, {Frinchaboy},
  {Allende Prieto}, {Barkhouser}, {Bizyaev}, {Blank}, {Brunner}, {Burton},
  {Carrera}, {Chojnowski}, {Cunha}, {Epstein}, {Fitzgerald}, {Garc{\'\i}a
  P{\'e}rez}, {Hearty}, {Henderson}, {Holtzman}, {Johnson}, {Lam}, {Lawler},
  {Maseman}, {M{\'e}sz{\'a}ros}, {Nelson}, {Nguyen}, {Nidever}, {Pinsonneault},
  {Shetrone}, {Smee}, {Smith}, {Stolberg}, {Skrutskie}, {Walker}, {Wilson},
  {Zasowski}, {Anders}, {Basu}, {Beland}, {Blanton}, {Bovy}, {Brownstein},
  {Carlberg}, {Chaplin}, {Chiappini}, {Eisenstein}, {Elsworth}, {Feuillet},
  {Fleming}, {Galbraith-Frew}, {Garc{\'\i}a}, {Garc{\'\i}a-Hern{\'a}ndez},
  {Gillespie}, {Girardi}, {Gunn}, {Hasselquist}, {Hayden}, {Hekker}, {Ivans},
  {Kinemuchi}, {Klaene}, {Mahadevan}, {Mathur}, {Mosser}, {Muna}, {Munn},
  {Nichol}, {O'Connell}, {Parejko}, {Robin}, {Rocha-Pinto}, {Schultheis},
  {Serenelli}, {Shane}, {Silva Aguirre}, {Sobeck}, {Thompson}, {Troup},
  {Weinberg}, \& {Zamora}}]{Majewski2017}
{Majewski}, S.~R., {Schiavon}, R.~P., {Frinchaboy}, P.~M., {et~al.} 2017, \aj,
  154, 94

\bibitem[{{Malhan} \& {Ibata}(2018)}]{Malhan2018}
{Malhan}, K., \& {Ibata}, R.~A. 2018, \mnras, 477, 4063

\bibitem[{{Martin} {et~al.}(2006){Martin}, {Irwin}, {Ibata}, {Conn}, {Lewis},
  {Bellazzini}, {Chapman}, \& {Tanvir}}]{Martin2006}
{Martin}, N.~F., {Irwin}, M.~J., {Ibata}, R.~A., {et~al.} 2006, \mnras, 367,
  L69

\bibitem[{{Mateu}(2023)}]{Mateu2023}
{Mateu}, C. 2023, \mnras, 520, 5225

\bibitem[{{McClure-Griffiths} {et~al.}(2009){McClure-Griffiths}, {Pisano},
  {Calabretta}, {Ford}, {Lockman}, {Staveley-Smith}, {Kalberla}, {Bailin},
  {Dedes}, {Janowiecki}, {Gibson}, {Murphy}, {Nakanishi}, \&
  {Newton-McGee}}]{McClure-Griffiths2009}
{McClure-Griffiths}, N.~M., {Pisano}, D.~J., {Calabretta}, M.~R., {et~al.}
  2009, \apjs, 181, 398

\bibitem[{{Morganson} {et~al.}(2016){Morganson}, {Conn}, {Rix}, {Bell},
  {Burgett}, {Chambers}, {Dolphin}, {Draper}, {Flewelling}, {Hodapp}, {Kaiser},
  {Magnier}, {Martin}, {Martinez-Delgado}, {Metcalfe}, {Schlafly}, {Slater},
  {Wainscoat}, \& {Waters}}]{Morganson2016}
{Morganson}, E., {Conn}, B., {Rix}, H.-W., {et~al.} 2016, \apj, 825, 140

\bibitem[{{Newberg} \& {Carlin}(2016)}]{Newberg2016}
{Newberg}, H.~J., \& {Carlin}, J. 2016, in Astrophysics and Space Science
  Library, Vol. 420, Tidal Streams in the Local Group and Beyond, ed. H.~J.
  {Newberg} \& J.~L. {Carlin}, 1

\bibitem[{{Newberg} {et~al.}(2002){Newberg}, {Yanny}, {Rockosi}, {Grebel},
  {Rix}, {Brinkmann}, {Csabai}, {Hennessy}, {Hindsley}, {Ibata}, {Ivezi{\'c}},
  {Lamb}, {Nash}, {Odenkirchen}, {Rave}, {Schneider}, {Smith}, {Stolte}, \&
  {York}}]{Newberg2002}
{Newberg}, H.~J., {Yanny}, B., {Rockosi}, C., {et~al.} 2002, \apj, 569, 245

\bibitem[{{Nidever} {et~al.}(2008){Nidever}, {Majewski}, \& {Butler
  Burton}}]{Nidever2008}
{Nidever}, D.~L., {Majewski}, S.~R., \& {Butler Burton}, W. 2008, \apj, 679,
  432

\bibitem[{{Nidever} {et~al.}(2019){Nidever}, {Price-Whelan}, {Choi}, {Beaton},
  {Hansen}, {Boubert}, {Aguado}, {Ezzeddine}, {Oh}, \& {Evans}}]{Nidever2019}
{Nidever}, D.~L., {Price-Whelan}, A.~M., {Choi}, Y., {et~al.} 2019, \apj, 887,
  115

\bibitem[{{Peebles}(1965)}]{Peebles1965}
{Peebles}, P.~J.~E. 1965, \apj, 142, 1317

\bibitem[{{Pieres} {et~al.}(2017){Pieres}, {Santiago}, {Drlica-Wagner},
  {Bechtol}, {Marel}, {Besla}, {Martin}, {Belokurov}, {Gallart},
  {Martinez-Delgado}, {Marshall}, {N{\"o}el}, {Majewski}, {Cioni}, {Li},
  {Hartley}, {Luque}, {Conn}, {Walker}, {Balbinot}, {Stringfellow}, {Olsen},
  {Nidever}, {da Costa}, {Ogando}, {Maia}, {Neto}, {Abbott}, {Abdalla},
  {Allam}, {Annis}, {Benoit-L{\'e}vy}, {Rosell}, {Kind}, {Carretero}, {Cunha},
  {D'Andrea}, {Desai}, {Diehl}, {Doel}, {Flaugher}, {Fosalba},
  {Garc{\'\i}a-Bellido}, {Gruen}, {Gruendl}, {Gschwend}, {Gutierrez},
  {Honscheid}, {James}, {Kuehn}, {Kuropatkin}, {Menanteau}, {Miquel}, {Plazas},
  {Romer}, {Sako}, {Sanchez}, {Scarpine}, {Schubnell}, {Sevilla-Noarbe},
  {Smith}, {Soares-Santos}, {Sobreira}, {Suchyta}, {Swanson}, {Tarle},
  {Tucker}, \& {Wester}}]{Pieres2017}
{Pieres}, A., {Santiago}, B.~X., {Drlica-Wagner}, A., {et~al.} 2017, \mnras,
  468, 1349

\bibitem[{{Press} \& {Schechter}(1974)}]{Press1974}
{Press}, W.~H., \& {Schechter}, P. 1974, \apj, 187, 425

\bibitem[{{Price-Whelan} {et~al.}(2019){Price-Whelan}, {Nidever}, {Choi},
  {Schlafly}, {Morton}, {Koposov}, \& {Belokurov}}]{PriceWhelan2019}
{Price-Whelan}, A.~M., {Nidever}, D.~L., {Choi}, Y., {et~al.} 2019, \apj, 887,
  19

\bibitem[{{Shipp} {et~al.}(2018){Shipp}, {Drlica-Wagner}, {Balbinot},
  {Ferguson}, {Erkal}, {Li}, {Bechtol}, {Belokurov}, {Buncher}, {Carollo},
  {Carrasco Kind}, {Kuehn}, {Marshall}, {Pace}, {Rykoff}, {Sevilla-Noarbe},
  {Sheldon}, {Strigari}, {Vivas}, {Yanny}, {Zenteno}, {Abbott}, {Abdalla},
  {Allam}, {Avila}, {Bertin}, {Brooks}, {Burke}, {Carretero}, {Castander},
  {Cawthon}, {Crocce}, {Cunha}, {D'Andrea}, {da Costa}, {Davis}, {De Vicente},
  {Desai}, {Diehl}, {Doel}, {Evrard}, {Flaugher}, {Fosalba}, {Frieman},
  {Garc{\'\i}a-Bellido}, {Gaztanaga}, {Gerdes}, {Gruen}, {Gruendl}, {Gschwend},
  {Gutierrez}, {Hartley}, {Honscheid}, {Hoyle}, {James}, {Johnson}, {Krause},
  {Kuropatkin}, {Lahav}, {Lin}, {Maia}, {March}, {Martini}, {Menanteau},
  {Miller}, {Miquel}, {Nichol}, {Plazas}, {Romer}, {Sako}, {Sanchez},
  {Santiago}, {Scarpine}, {Schindler}, {Schubnell}, {Smith}, {Smith},
  {Sobreira}, {Suchyta}, {Swanson}, {Tarle}, {Thomas}, {Tucker}, {Walker},
  {Wechsler}, \& {DES Collaboration}}]{shipp2018}
{Shipp}, N., {Drlica-Wagner}, A., {Balbinot}, E., {et~al.} 2018, \apj, 862, 114

\bibitem[{{Slater} {et~al.}(2014){Slater}, {Bell}, {Schlafly}, {Morganson},
  {Martin}, {Rix}, {Pe{\~n}arrubia}, {Bernard}, {Ferguson}, {Martinez-Delgado},
  {Wyse}, {Burgett}, {Chambers}, {Draper}, {Hodapp}, {Kaiser}, {Magnier},
  {Metcalfe}, {Price}, {Tonry}, {Wainscoat}, \& {Waters}}]{Slater2014}
{Slater}, C.~T., {Bell}, E.~F., {Schlafly}, E.~F., {et~al.} 2014, \apj, 791, 9

\bibitem[{{Tonry} {et~al.}(2018){Tonry}, {Denneau}, {Heinze}, {Stalder},
  {Smith}, {Smartt}, {Stubbs}, {Weiland}, \& {Rest}}]{Tonry2018}
{Tonry}, J.~L., {Denneau}, L., {Heinze}, A.~N., {et~al.} 2018, \pasp, 130,
  064505

\bibitem[{Virtanen {et~al.}(2020)Virtanen, Gommers, Oliphant, Haberland, Reddy,
  Cournapeau, Burovski, Peterson, Weckesser, Bright, {van der Walt}, Brett,
  Wilson, Millman, Mayorov, Nelson, Jones, Kern, Larson, Carey, Polat, Feng,
  Moore, {VanderPlas}, Laxalde, Perktold, Cimrman, Henriksen, Quintero, Harris,
  Archibald, Ribeiro, Pedregosa, {van Mulbregt}, \& {SciPy 1.0
  Contributors}}]{virtanen2020scipy}
Virtanen, P., Gommers, R., Oliphant, T.~E., {et~al.} 2020, Nature Methods, 17,
  261

\bibitem[{York(2000)}]{York2000}
York, D.~G. 2000, AJ, 120, 9.
\newblock \url{http://arxiv.org/abs/astro-ph/0006396}

\bibitem[{{Zhang} {et~al.}(2023){Zhang}, {Green}, \& {Rix}}]{Zhang2023}
{Zhang}, X., {Green}, G.~M., \& {Rix}, H.-W. 2023, \mnras, arXiv:2303.03420

\end{thebibliography}



\end{document}